** This is a preprint version and has not been peer-reviewed**

# Machine learning-based multimodal prognostic models integrating pathology images and high-throughput omic data for overall survival prediction in cancer: a systematic review

**Authors:**

Charlotte Jennings[1,2], Andrew Broad[1,2], Lucy Godson[1,2], Emily Clarke[1,2], David Westhead[2], Darren Treanor[1,2,3].

**Affiliations:**

1. Leeds Teaching Hospitals NHS Trust, Leeds, UK
2. University of Leeds, Leeds, UK
3. Department of Clinical Pathology, Linköping University, Sweden

**Correspondence:**

Dr Charlotte Jennings

Email: charlotte.jennings1@nhs.net;

# Abstract

**Background:**

Multimodal machine learning models that integrate histopathology and molecular data offer a promising approach to cancer prognostication and are an area of rapid research growth. However, the methodological quality, reporting standards, and clinical relevance of these models remain unclear. We conducted the first systematic review of studies in which machine learning or deep learning methods were used to combine pathology whole slide images and high-throughput omic data to predict overall survival in cancer.

**Methods:**

Studies were identified by a systematic search of EMBASE, PubMed and Cochrane CENTRAL, and were conducted up to 12th August 2024, with citation searching for additional studies. Data were extracted using the CHARMS checklist and risk of bias assessment performed using the PROBAST+AI tool. A narrative synthesis approach was conducted following the SWiM and PRISMA 2020 reporting guidelines. The study was registered on PROSPERO (CRD42024594745).

**Results:**

Forty-eight studies were eligible for inclusion, all published since 2017. Studies covered survival prediction of cancer across 19 organs. The Cancer Genome Atlas dataset was used in all studies. The modelling approaches ranged from regularised Cox regression methods (n=4) to classical machine learning methods (n=13) and deep learning approaches (n=31). Model performances, as measured by the concordance index (c-index), ranged from 0.550 – 0.857 and, where comparison was available with simpler unimodal models, the multimodal models outperformed them in all but one study. All studies were judged to be at unclear or high overall risk of bias, with inconsistent reporting and limited external validation. Consideration of clinical utility was infrequent.

**Conclusion:**

Multimodal survival prediction using WSI and omics is a fast-growing field with early promising results. However, this review shows that studies suffer from methodological bias, narrow data sources, and poor clinical contextualisation. Progress will require greater focus on robust reporting, diverse datasets, and evaluation of real-world clinical utility.

**Funding:**

This work was funded by the National Pathology Imaging Cooperative (NPIC), Leeds Teaching Hospitals NHS Trust, UK. NPIC (Project no. 104687) is supported by the Data to Early Diagnosis and Precision Medicine strand of the government's Industrial Strategy Challenge Fund, managed and delivered by UK Research and Innovation.

## 1.1 Introduction

Cancer remains a key health priority globally, with an estimated 18.1 million new cases of cancer worldwide in 2020 and a projected increase to an annual incidence of 28 million by 2040.[1] Prognostic models that accurately stratify patients by survival risk are essential for guiding treatment decisions, clinical trial design, and for health resource planning making them valuable to patients, clinicians, researchers and policy makers. [2,3]

Routine cancer care involves the integration of multi-modal data—including clinical, radiological, pathological, and molecular inputs—to guide decision making. Digitisation of pathology slides into whole slide images (WSI) and advances in omics now enable the computational analysis of these data, supporting precision oncology. Clinical prediction models are tools that estimate a patient's risk of an outcome by combining multiple covariates, and have traditionally relied on regression-based methods. With the increasing scale and complexity of multi-modal cancer data available, there has been increasing interest in applying machine learning (ML) and deep learning (DL) approaches to build more flexible, data-driven prognostic models.[4]

ML models have already demonstrated improved survival prediction performance in single-modality settings, including WSI data[5,6], and multiomics.[7] More recent studies report enhanced prognostic predictions by the integration of these data, assumed to be a result of capturing complementary biological signals. To date, clinical translation of ML approaches in other domains has been very low.[8] There is a need to understand the maturity of this field, whether there is the potential for real-world utility and if the performance gain of these models is sufficient to justify the high costs of laboratory data generation and computation.

### Related work

A systematic review by Schneider *et al* (2022) evaluated studies integrating pathology image and genomic data by DL (searches up to June 2021).[9] This review identified 11 studies, including seven predictive of survival (4 cancer-specific survival, 3 overall survival) and four related to other tasks, including; prediction of microsatellite instability (MSI) status, malignant versus benign differentiation and subtype prediction. The multimodal approach was shown to be superior to unimodal methods in all studies, however the heterogeneity of studies limited further conclusions. A more general review by Cui *et al* (2023) across the medical domain also determined that multimodal DL prediction models typically surpass unimodal models for disease diagnosis and prognosis tasks.[10] This selective review evaluated thirty-four studies using feature-level multimodal DL-based fusion to integrate image and non-image data, and outlined emerging multimodal frameworks and fusion approaches. Data availability and lack of explainability were identified as key limitations to these studies.

This review provides the first focussed review on prognostic prediction in the cancer domain through high-throughput omic and WSI integration. In line with best practice methodology, we systematically review all literature in which these data are combined as predictors for the prediction of overall survival (OS) in cancer patients using ML or DL techniques. We focused on OS as a universally applicable, clinically meaningful outcome that is consistently reported across cancer types and public datasets, enabling standardised comparison across studies. While OS reflects factors beyond tumour biology, its broad availability and relevance made it the most practical and informative endpoint for this review.

**Objectives**

This review specifically addresses the following questions;

- What is the prediction performance for OS?
- What cancer domains have these techniques been applied to?
- What types and scale of data sources are used?
- What methods are used for data processing and integration of data?
- What is the quality of these studies and their reporting?

## 1.2 Methods

This systematic review without meta-analysis was conducted in accordance with the guidelines for "Preferred Reporting Items for Systematic Reviews and Meta-Analyses" (PRISMA).[11] Due to the highly heterogenous nature of included studies a narrative synthesis approach was used in line with the "Synthesis without meta-analysis" (SWiM) guidelines.[12] The protocol for this review is available at https://www.crd.york.ac.uk/PROSPERO/display_record.php?RecordID=594745 and was approved before the search results were screened for inclusion (Registration CRD42024594745).

### 1.2.1 Eligibility criteria

Primary, peer-reviewed research studies reporting the development and/or validation of a prediction model developed using ML methods and integrating pathology WSI data with high-throughput omic data were sought. Only peer-reviewed journal or conference papers were included.

The population included human participants with any cancer diagnosis. The outcome of interest was OS predicted at any time point. Predictive models for other outcomes, including cancer-specific survival, were excluded. The search strategy was geared towards selection of typical ML or DL approaches. However, to recognise ML methodologies which build on traditional statistical models, studies using regression approaches were accepted as ML models if they were defined as such by the authors of a study – an approach used previously [8,13]. Studies focusing solely on prognostic factor identification or lacking image and omic data integration were excluded. Only conventional light microscopy imaging of surgical pathology specimens was included, and studies relating to cytology, forensic or other material were not included, nor were studies using other imaging techniques. Studies in which all models also integrated additional radiology imaging were not included.

Studies not in English were excluded due to limitations of the researchers. No exclusions were applied for date of publication.

### 1.2.2 Data sources and search strategy

Searches were conducted in two research databases (EMBASE and PubMed) and one trial registry (Cochrane Central Register of Controlled Trials (CENTRAL)) from inception to 12th August 2024. Searches were restricted to human studies and English language.

The search strategy was composed of terms related to pathology whole slide images, omic data, machine learning, and cancer. For each category multiple terms were combined with the *OR* operator, before combining categories with the *AND* operator. The full search strategy is available in supplementary material. Citation checking was also conducted (October 2024) using the selected studies from the abstract screening processes as seed references in a forward and back seed approach using *citationchaser shiny app*[14,15].

### 1.2.3 Study selection

All studies from the literature searches were imported to the *Rayyan* software to manage the screening stages of the review across the team[16]. First, duplicates were manually removed by C.J. supported by the duplicate detection tool within *Rayyan*. Studies were then screened according to predefined algorithms. One investigator (C.J.) screened all titles and abstracts, with a second independent screen performed by either E.C. or L.G.. Disagreements were resolved by discussion with the third investigator. Full text articles were screened in the same way. The screening algorithms are available in the supplementary material.

### 1.2.4 Data extraction

Data extraction for each study was performed independently by two reviewers using a predefined data extraction spreadsheet, which was adapted from a previously developed template[17]. The template was designed with reference to the “Critical Appraisal and Data Extraction for Systematic Reviews of Prediction Modelling Studies” (CHARMS) checklist and “Prediction model Risk OF Bias ASsessment Tool” (PROBAST), and was updated to align with PROBAST+AI on its release[18–20]. C.J. reviewed all studies, while the second independent review was performed by either A.B. or L.G.. Disagreements were resolved by discussion with the third investigator.

Study level information extracted from papers, included: study demographics, cancer domain, inclusion and exclusion criteria, dataset details and participant characteristics. Models meeting the inclusion criteria within these studies were appraised. Detailed information was extracted as below for the best performing OS prediction model in each study meeting the inclusion criteria (defined as our model of interest), with information about any additional models in the paper extracted in a summarised format. Model specific data fields included: approaches to feature generation and selection, model architecture and fusion approaches, and performance measures. Several fields were added or clarified during data extraction with the agreement of all researchers. Any changes were retroactively applied to all previously extracted studies. The final data extraction template is summarised in supplementary material.

Information was sought from the full-text articles, as well as supplementary materials where appropriate. Inferences were only made where both researchers were confident to do so and labelled as *unclear* where this was not possible. The well-known nature of the heavily used The Cancer Genome Atlas (TCGA) dataset meant we were able to deduce some characteristics of the data used even when not explicitly provided. For example, TCGA only has H&E images and refers to slides which underwent FFPE fixation as “Diagnostic slides” and frozen sections as “Tissue slides”. In these situations, assumed data is marked by an asterisk to help understand completeness of reporting of these fields.

Authors were not contacted for clarification where information was missing or unclear.

#### 1.2.4.1 Risk of bias assessment

The consensus developed PROBAST+AI tool was used to assess the models of interest in this study.[20] The tool assesses the likelihood of results being affected by the study design, conduct or analysis. The model development and evaluation process are assessed by signalling questions across four domains (participants, predictors, outcome and analysis), which probe issues including appropriate cohort selection, predictor definition and measurement, outcome clarity, handling of missing data, overfitting and validation methods. Through these questions reviewers are guided towards judging whether each domain is of high, low or unclear concern for quality and risk of bias. Unclear ratings are used where there is not enough information to make a full assessment. The overall quality and risk of bias ratings are determined by the worst domain score. An overall low-concern rating requires low scores in every domain, whereas a high-concern rating in a single domain will lead to high-concern overall.

The PROBAST+AI tool also assesses the applicability of prediction models to the specific criteria of the systematic review across the predictor, participant and outcome domains. Studies are also determined to be of low concern or high concern, or unclear where there is not enough information to make a full assessment.

Owing to the potential subjective interpretation of the signalling questions in the PROBAST+AI tool, two independent researchers completed this process for each model, with disagreements resolved as previous by the third. Each paper was assessed by a pathologist and a computer scientist.

### 1.2.5 Data synthesis

The results of the literature search and screening process were summarised in a PRISMA flow chart generated by *PRISMA Flow Diagram Shiny app*.[21] All extracted data were summarised in two tables, covering study level and model characteristics.

The data synthesis did not include any meta-analysis due to the diversity of the methods used and was designed with reference to the SWiM protocol.[12] For model characteristics, studies were grouped by the modelling method used and further ordered by year of publication to enable appreciation of any method shift over time. The concordance-index (c-index) was chosen as the comparator metric between studies because it was anticipated to be the most consistently reported evaluation metric based on preliminary searches. Two graphs were generated using the c-index, where it was available, to show a) comparison of performance between the multimodal model of interest and unimodal model performances explored in the study, and b) performance variation across different cancer types. Area under the curve (AUC) and time-specific AUC values were also extracted where available. Summary metrics used for calibration, overall performance and clinical utility were also tabulated.

The results of the PROBAST+AI assessments were tabulated by study and presented as summary graphs for quality, risk of bias and applicability.

## 1.3 Results

Literature searches returned a total of 457 studies, of which 132 were duplicates. Two hundred and eighty-four records were excluded during abstract and title screening and 21 were excluded by the full text screening process. An additional 1840 titles and abstracts were screened as part of citation searches, of which 36 new studies were screened in full text and 29 included in the review Figure 1. The study characteristics are shown in Table 1. Details of the models evaluated are shown in Table 2.

Of the 48 included studies[22–69], ten of these were conference papers with the remaining 38 being journal papers. All included studies were published since 2017. Studies appeared in 31 different publications, mainly in computer science and computational biology focussed publications (n=22), but also in scientific, medical and medical imaging journals. Eight of the studies were in cancer specific journals. Lead authors were distributed across 5 countries; China (n=33), USA (n=10), Germany (n=2), Hong Kong (n=2) and Spain (n=1).

Important studies in this field but not eligible for this review, include Mobadersany et al[70] who integrated IDH1 mutation status (*not high-throughput omic*) with image data for survival prediction in glioma, studies in which a model was developed to predict molecular status from the WSI and then fed into an image-based prediction model (*no data integration)*[71,72], and studies where all developed models incorporated radiology data alongside WSI and omics.[73,74]

### 1.3.1 Risk of bias assessments

The results of the PROBAST+AI assessments for each model of interest are shown in Table 3 and risk of bias and applicability across the review are summarised in Figure 2. Due to the limited use of external datasets for validation, assessments across the participant, predictor and outcome domains for quality of model development and risk of bias of evaluation are usually the same.

**Participant**

All studies were had an unclear (n=47) or high concern (n=1) of bias in the participant domain. This was predominantly due to a lack of information provided about the TCGA datasets used, including recruitment methods, setting, and inclusion and exclusion criteria. Only a couple of studies included additional detail, such as recruitment dates[24] and number of sites[33]. Where additional datasets were used, a greater level of information was usually provided, but this still provided an incomplete picture of the source of data in all studies.[22,39,42,48,59,65,67]

Individual study criteria for inclusion or exclusion, where stated, were usually data based; defined by presence in a given TCGA data subset, by data quality and/or by completeness of matched data across modalities. Only two studies had clinically orientated inclusion criteria.[46,54] Presentation of participant characteristics was poor, with 18 studies (38%) providing no description of the cohort used, see Table 1. The remaining studies presented variable coverage of participant demographics, disease and treatment information. In one study, this was presented comprehensively enough to assess the representativeness of the dataset.[54] In this study of oral squamous cell carcinoma, participant demographics were broadly representative of the USA population[75], however were skewed towards a population with a higher pathologic stage of disease.

### Predictor

Studies generally provided no information about data acquisition processes making it difficult to assess similarity of predictor assessment. Within the TCGA dataset, there are known to be differences in protocols and processes between contributing institutions which are sufficient to cause batch effect in both sequencing and imaging data.[76–80] Whilst the variation was not well described, a few of the papers outlined pre-processing steps in one or both modalities designed to address these issues. Only one study described performing these across both modalities and was rated as low concern.[48]

In one of the studies, participants from one contributing site in the TCGA dataset were removed after persistent variation in stain post processing, which raised concern for selection bias.[67] Another study integrated omic data from different datasets, generated by different methods without clear discussion about the comparability of these methods and any consequent differences when used to infer the homologous recombination (HRD) status from them.[22]

### Outcome

Assessments of quality and RoB were generally low due to nature of the outcome (death), which is not so sensitive to bias. However, studies which did not provide enough information to appreciate the intended outcome prediction time horizon or assess the length of follow up of the patients used in the study were rated as unclear.[26,29,33]

### Analysis

In this domain, there was high concern of quality in 25 studies and of risk of bias in 45 studies. Common factors contributed to these decisions as outlined below.

Assessment of adequacy of sample size in ML and DL approaches is difficult due to the influence of regularisation techniques, shrinkage methods, and hyperparameter tuning strategies aimed to prevent overfitting. The application of the traditional “rule of thumb” 10 events per variable are inappropriate in this setting and there are no clearly defined minimum criteria. As shown by Collins et al. (2021), model complexity and tuning parameters substantially increase data demands, and the number of outcome events may be more influential than total sample size alone.[81] The use of cross validation, which increases the effective size of the test set thereby reducing over optimism in the results further complicates assessment. Recent guidance encourages authors to justify their sample size for the reader.[82] However, none of the studies presented any calculation or justification of their sample size and many were missing key data such as sample size (n=1), number of events (n=26), number of candidate (n=24) or final predictors (n=9) used in the models making it difficult to assess event per variable.

Another key consideration was the common use of a complete case analysis approach, with subjects with missing data modalities excluded and variables with missing values in included patients often removed. This approach can introduce bias if the missingness is not at random. However, it is also appreciated to make sense as part of predictor reduction strategy. Three studies described a proactive approach to data missingness and are discussed in the model section below.

All studies reported measures of discrimination, which test the model's ability to correctly distinguish between patients who experience the event (e.g., death) earlier versus later. However, only four evaluated calibration, which refers to how closely a survival prediction model's estimated probabilities of survival match the observed survival outcomes. These are complementary aspects of model performance and evaluating only one can miss serious model flaws—e.g., a model may rank patients correctly (good discrimination) but still systematically over- or under-predict risk (poor calibration), leading to biased clinical decisions.

Additionally, the presentation of results did not include metrics of variability in six studies and in a further 19 the metric used was unclear.

### Overall quality and risk of bias

None of the studies had high or unclear ratings in just a single domain, indicating literature in this area faces broad methodological and reporting issues.

### Overall applicability

This systematic review was designed to be broad in scope. All studies were assessed as low concern in the participant domain as all were studies of cancer patients. In the predictor domain, all studies were rated as low concern, clearly using high-throughput omic data and pathology whole slide images in their models. In the outcome domain, all studies were predictive of survival. However, 31 studies were not explicit about an outcome of OS and used ambiguous phrasing such as “cancer survival” or just “survival”. The most used dataset in this study (TCGA) is known to have associated cancer specific and OS data and as such specific phrasing was sought or the applicability of the study was assessed as unclear.[83]

### 1.3.2 Data synthesis results

#### Datasets

Twenty-nine studies focused on a cancer of a single organ/anatomical region, while 19 were multi-tissue studies. Nineteen different organs were studied, with the most studied being brain, breast, lung and kidney. Figure 3 summarises frequencies of the cancer domains studied.

Alongside WSIs, gene expression (mRNA) data was used in all studies, with additional omics used in variable combinations; somatic mutation data (n=9), micro-RNA (n= 5), copy number variation (CNV) (n=16), single nucleotide variation (SNV) (n = 3), DNA methylation (n = 5) and protein expression (n = 5). Clinical data was integrated as an additional modality into 15 models and the variables used most commonly were age, gender and stage. The number of participants used per model of interest ranged from 63 to 11160.

The TCGA dataset was used in some capacity in all studies, 47 for training, and in three studies as an independent external dataset. TCGA is well known, multi-institutional public dataset comprising mixed omic data and WSIs for over 11,000 participants from the USA spanning 33 cancer types.

Other datasets used in multimodal model development and/or validation were from Clinical Proteomic Tumour Analysis Consortium (CPTAC)[48,65], Memorial Sloan Kettering Cancer Centre[22], Molecular Taxonomy of Breast Cancer International Consortium (METABRIC)[42], NCT-CRC-HE-100[39], Paediatric Brain Tumour Atlas[48], National Lung Screening Trial[65], Tumour infiltrating lymphocytes in breast cancer (TIGER)[42], Ruijin Hospital (Shanghai)[59] and Wayne State University[67]. The countries from which data were used include USA, China, Belgium and Netherlands. The predominant usage of open access public data repositories meant data from these studies was largely assessed as available, however only 18 studies included specific statements relating to data availability

– asterisks in Table 1 indicate where data availability was assumed. Those that did not, all exclusively used the TCGA dataset.

## Modelling methods

In this review, the model of interest was defined as the best performing model. Many of the studies, created more than one model by either: independently training and evaluating the same model architecture on datasets of multiple cancer types, combining different omic combinations with image and/or clinical data into the model, or by testing multiple different model architectures or fusion approaches. Where additional models were present in a study, this is indicated in Table S1.

### *Models*

Modelling approaches included cox regression models with ML-based regularisation and/or feature selection (n=4)[28,44–46], classical ML methods (n=13) and DL methods (n=31). ML methods included seven models based on random forest [24,25,32,34,54,60,61], a multi constraint latent representation framework[41], two variants of canonical correlation analysis[49,50], two multiple kernel learning models[51,63] and a multiview subspace learning model[64]. Of the 31 DL models, 25 used a cox-based loss as the objective function for survival prediction (log partial or negative log partial likelihood), but cross entropy loss[26,65] and binary cross entropy loss[37,55] were also used. One study used a custom likelihood and ranking loss function (DeepHit).[33]

Within the models, fusion of data at a feature level was more common (n=42) than decision level fusion[22,31,48]. Three studies used a hybrid approach, processing one modality to a decision level (risk score) for integration with features of another modality in the prediction model.[30,32,57] Studies using feature fusion methods leveraged mechanisms to capture cross-modal interactions, most commonly through attention-based approaches, where models compare different parts of the input, assign relevance scores, and weight them to focus on the most important information (see Table S1).

While most models required fully matched datasets, four studies implemented explicit strategies to accommodate missing data modalities without excluding patients.[23,31,43,53] Qiu et al. (2024) and Hou et al. (2023) used latent-space reconstruction—via variational autoencoders and online masked autoencoders, respectively—to recover missing features.[31,43] In contrast, Vale-Silva and Rohr (2021) used multimodal dropout during training and standard imputation for partial missingness, while Cheerla and Gevaert (2019) dynamically reweighted available modalities during fusion.[23,53] Despite differing mechanisms, all approaches aimed to maintain performance under incomplete input and reported reduced accuracy when modalities were missing. Models lacking such strategies may be vulnerable to overfitting and underperformance in real-world, incomplete datasets.

Model code was described as available in 25 studies, partially available code (as "related core code") in one study[41] and not commented on in the remaining 22 studies. Using links to code repositories, in four studies code could not be found[28], was described as "coming soon" [68] or had been taken down and was now only available on request.[41,64]

*Data pre-processing*

These complex data require pre-processing before they can be used in a model and reduction in the dimension of the also data helps to reduce model overfitting. Such steps were variably described in these studies and are summarised below and in supplementary table S1.

Most researchers performed tissue segmentation and/or stain normalisation of WSIs before dividing the original WSI into patches, ranging from 16x16 to 5000x5000 pixels, most commonly 1000x1000 pixels (n=12) or 512x512 pixels (n=11). Four studies used multiscale patches.[27,38,39,56] Further selection of patches for inclusion was commented on in 32 studies and was performed at random (n=9), by RGB density (n=8), nuclear density (n=2), proximity to RGB mean (n=1), based on expert defined regions of interest (n=9), by exclusion of patches with low tissue coverage (n=2) or for not meeting feature extraction model criteria (n=1). Image features were generated equally by hand-crafted (n=23) or learned approaches (n=23) with two studies using a combined approach. CellProfiler was the most popular tool for generating hand-crafted features (n=17) and ResNet50 the most popular architecture for learned features (n=7). Some models incorporated graph-based ML[49,50], graph neural network,[31,65] transformer[38,39,56] or foundation model[69] approaches for feature generation.

Pre-processing of omic data, where described, included removal of missing data, normalisation of read counts, and discretisation of expression or CNV data into categorical variables. The two studies which described correction for batch effect, both used ComBATSeq.[48,65] In some studies, omic data was used directly, however several performed feature generation, including; knowledge-based approaches, such as gene set enrichment analysis (n=1) and inference of known mutational signatures (n=2), gene modularisation by data driven methods, including Weighted Gene Co-expression Network Analysis (n=6) and local maximal Quasi-Clique Merger (n=4), and generation of feature representations by ML methods (n=15).

Steps to reduce number of predictors were described in most studies, see Supplementary Table S1, and used a combination of data and knowledge-based approaches. The heterogeneity of approaches to feature generation (predictors) makes it difficult to compare number of predictors across studies. It was also challenging to determine the predictor number in many of the models. Particularly those where features are generated via DL and have cross-modal interactions resulting in complex representations from the WSI and omic inputs.

**Model performance and evaluation**

All studies performed some form of internal validation, with most using a cross-validation approach (n=44) and the remaining few using random data splitting (n=3) and bootstrapping (n=1).

As discussed, studies focussed on measures of discrimination for reporting model performance. The c-index was the most commonly reported metric (39/48). It is a measure of the ability of the model to rank patients in terms of their predicted risk of event by indicating the proportion of all usable patient pairs in which the patient with the shorter observed survival time also has the higher predicted risk. A c-index of 1 indicates perfect performance and 0.5 is equivalent to random predictions. Results across these studies ranged from 0.550-0.857. Studies which additionally modelled survival as a binary classification task (i.e.. predicting longer or shorter survival based on defined thresholds) reported AUC metrics (n=14). An AUC of 1 reflects perfect classification and 0.5 random guessing. Results across studies ranged 0.662-0.932. Time-specific AUC, which measures how well the model can distinguish between classes at a set time point, was reported in eight studies with results across multiple time points but most commonly for 1-, 3- and 5-year survival.

Two studies performed neither of these analyses, instead assessing model performance by hazard ratios[42] and F1 score[49].

Five studies performed external validation on their multimodal model[48,59,64,65,67], with an additional four studies undertaking external validation but only of unimodal models in their study.[24,25,57,61] Results of multimodal external validation ranged from 0.583-0.778 (c-index).

Only seven studies performed clinical utility analyses, all via decision curve analysis. [24,25,30,32,34,60,61] Four additionally explored end-user presentation of the model via development of nomograms[24,30,32,34] and a further study generated an online tool.[57]

*Unimodal performance comparison*

Thirty-four out of the 48 studies made a performance comparison between the multimodal model of interest and unimodal models based on clinical, omic or image data. Where the c-index was available as a metric (n = 25), these have been plotted in Figure 4. Studies which made a unimodal comparison but assessed this by another metric (AUC = 10, F1 = 1) could not be included. In all but one study, multimodal models outperformed unimodal models but the extent of improvement in prediction is highly variable. Vale Silva and Rohr's MultiSurv model showed greater performance on clinical data (c-index 0.809) than with their multimodal model (c-index 0.787, multiomics+WSI), or image (c-index 0.569) and omic (c-index 0.758, mRNA) based models.[53]

*Other performance comparisons*

Performance across cancer types was compared where there were results presented in greater than five studies, see Figure S2. This synthesis used data from all datasets evaluated in the studies for the model of interest, including alternative dataset results from multi-cancer studies. In this descriptive comparison, variation in performance across different cancer types is observed. Studies of low-grade glioma, glioblastoma and renal papillary cell carcinoma are generally best performing so far.

The number of patients in the studies showed a weak trend toward increased performance with increased sample size. The presence of 400 patients or more appeared to be sufficient to achieve optimal results on internal test sets. Studies with fewer participants generally performed the worse (Figure S3). There was no meaningful association between performance and increasing event rate, although only 15 studies reported data sufficiently to be included in this analysis (Figure S4). These comparisons are illustrative only, due to heterogeneity in methods and risk of bias noted in this review.

## 1.4 Discussion

Our review highlights the rapid growth of this research domain in the last 5 years. Models have been developed across a wide spectrum of cancers, representing 19 organ systems and 23 specific diagnostic subtypes. Similar to previous works, we find that multimodal integration of WSI and omic data, generally leads to improved prognosis prediction. Despite methodological progress, these models are far from ready for clinical evaluation.

All included studies were judged to be at high or unclear risk of bias, highlighting significant issues in study design, reporting, and validation. As a result of the significant heterogeneity and potential bias in these models, we are not able to more rigorously assess these models by meta-analysis. However, the range in reported performance is wide (c-index, 0.550-0.857) and the gains over unimodal models, where evaluated, are often small. As such, we remain uncertain about the potential clinical value of these models for the future. ML methods are an attractive prospect, able to flexibly handle high-dimensional, non-linear data. However, despite these perceived advantages many studies of ML prediction models in other domains have found no additional benefit of ML over traditional models.[8] Given the expense of data generation, that is not part of routine clinical workflows (e.g. transcriptomic data), the gain would need to be substantial to be used clinically. Model complexity is another factor in the cost of implementing these multimodal models and only one study in this review evaluated compute requirements of their model.[35] Future studies must compare models against standard clinical prediction methods and begin to evaluate the clinical utility and cost benefit of their models, and not just seek to outperform the discrimination metrics of competing models.

These shortcomings seem reflective of the immaturity of this field in which methodological exploration for handling complex image data and innovation of ML approaches in predictive modelling has been the focus so far. DL-based models predominate in these studies, as does feature fusion of data. Some of the more recent DL models make use of graph networks, transformers and foundation models – keeping pace with DL developments in other fields. Whilst many studies develop approaches to capture cross-modal interactions and maximise the benefit of this complementary data, relatively fewer have explored models which are robust to missing data modalities – a common real-world scenario. Of concern, most studies lacked clinical contextualisation, with populations often defined by data availability rather than intended use. Important data issues, including assessment of the representativeness of the data for demographic and clinical features, and removal of missing data without consideration and evaluation of whether this is missing at random or not, were almost universally overlooked. The well-known nature of TCGA used by most studies may be leading authors to assume knowledge from readers. Yet, there are well-documented batch effects and participant shifts versus the general population for age, race and stage in this dataset, and more detail is important to the interpretation of these models.[76,77,79,84–87] Despite reliance on public datasets, open science practices were inconsistently adopted, with limited code and data availability described.

The heavy dependence on the TCGA dataset, used in all studies, may be an indicator that we don't yet have the necessary data to build and test these models. The whole field may be at risk of overfitting to the features of this single data set. The oncology community need to build more easily accessible datasets with detailed meta data and appropriate longitudinal follow up to provide clear and independent data sets on which to validate models.

### 1.4.1 Limitations of the review

The review protocol was designed to limit bias and maximise inclusion of eligible studies. The review was restricted to studies published in English and as such may have missed studies in other languages. All stages of screening and data extraction were performed by two independent researchers, except for the initial duplicate screening which was performed by a single researcher (with the aid of *Rayyan* software[16]), raising the possibility of incorrectly excluding studies in error.

The significant heterogeneity between the studies in this review precluded more detailed data synthesis and meta-analysis which limited the conclusions that can be made. Due to the rapid expansion of this research area, a future a review focussed on a specific cancer types or specific modelling methods may provide the opportunity for a more rigorous analysis.

Finally, this review focussed on models which directly integrated pathology and omic data, overlooking studies outside this domain. Most relevant for future consideration may be the evaluation of models which additionally incorporate radiology data [73,74], replicating the current clinical paradigm, and models which used omic and image data in a pipeline. Such approaches generally first generate an omic prediction from the image, then generate an image-based prediction model.[71,72] The rising availability and use of spatial omic data may propel this field forward, bridging the gap between morphology and bulk omic data for greater biological and interpretable insights.[88,89]

### 1.4.2 Current limitations and future recommendations

Our review finds that many aspects of methodology and reporting were suboptimal in these studies, in keeping with findings of other systematic reviews of prognostic models in the oncology domain.[3,90,91] Whilst there is rapid proliferation of prognostic models in cancer, the translation rate from academic generation to clinical implementation is very low.[8] The next phase of model development in this domain will benefit from reflections on lessons learnt about how to design clinically useful and usable models.[84] Future research in this area should adhere to reporting guidance, recently updated to better reflect complexities of ML-based modelling. TRIPOD+AI is a 27-item checklist designed to promote transparent reporting of studies developing, validating or updating prediction models.[92] The PROBAST+AI tool used in this review, also outlines best practice approaches to data selection and handling, modelling, and reporting.[93]

Even for well-developed and validated models, clinical implementation is not realistic in the near future. Multimodal models are being developed in advance of the widespread clinical availability of high-throughput omic and WSI data. In the UK, for example, approximately 10% of the molecular workload in the National Health Service is whole genome sequencing with the majority of the work centred on targeted mutation testing and smaller gene panels – there is no routine RNA sequencing.[94] In parallel, digitisation of pathology services across the UK is not yet complete.[95] Understanding of where and when these models may provide additional benefit to patients from well-conducted research may inform how such services develop.

Finally, research in this field will not progress without wider use and availability of additional datasets. So far, generation of these complex data has largely been the reserve of publicly funded initiatives. Other pan cancer projects such as the Applied Proteogenomic Organisational Learning Outcomes (APOLLO) network[96] and the Cancer Moonshot Biobank[97], are increasing the availability of publicly accessible omic data with matched pathology and radiological imaging data. Genomics England's 100,000 Genomes Project Multimodal Programme is a large apply to access dataset also on the horizon.[98] However, existing apply to access datasets, such as the Children's Brain Tumour Tissue Consortium[99] and METABRIC (breast)[100] seem under used in this field so far. This suggests there may be challenges for researchers in finding and accessing data, or that there is no funding or motivation to do so. Critical to clinical implementation will be increased availability and use of datasets from more diverse populations (genetic ancestry, geographic, environmental) and datasets which are representative of both the medical diversity of a condition, and the technical diversity of the data generated.[101]

### 1.4.3 Conclusion

This review of machine learning-based multimodal predictive models found 48 eligible studies predicting overall survival in cancer participants. This demonstrates significant growth in this research field since first works were identified in 2017. Progress has centred on methodological innovation, particularly through deep learning approaches. However, studies are limited by poor reporting, limited validation and lack of clinical contextualisation. To advance towards clinical impact, future research must prioritise transparent reporting, large diverse datasets, meaningful comparison to clinical standards, and a clear demonstration of how and where these models might be used in practice. Striking the right balance between model complexity, computational demands, and meaningful gains in predictive performance will be essential for real-world implementation—particularly in resource-constrained healthcare settings where cost-effectiveness is critical.

#### AUTHOR CONTRIBUTION STATEMENT

Authors C.J., D.T. and D.W. planned study. C.J. conducted the searches. Abstracts were screened by C.J., E.C., L.G. Full texts articles were screened by C.J., E.C., L.G.. Data extraction was performed by A.B., C.J., L.G.. C.J. analysed the data and wrote the manuscript, which was revised by A.B., D.T., D.W., E.C., L.G. All authors approved the manuscript for publication.

**FUNDING**

National Pathology Imaging Co-operative, NPIC (Project no. 104687) is supported by a £50m investment from the Data to Early Diagnosis and Precision Medicine strand of the government's Industrial Strategy Challenge Fund, managed and delivered by UK Research and Innovation (UKRI).

**DATA AVAILABILITY STATEMENT**

All data generated or analysed during this study are included in this published article and its supplementary information files. Extra data may be available from the corresponding author on request.

**COMPETING INTERESTS STATEMENT**

The authors declare no competing interests.

## 1.6 Figures

**Figure 1. PRISMA  2020 flow diagram**. PRISMA 2020 flowchart of the study identification and selection process for the systematic review. Records were screened on titles and abstracts, and reports were assessed based on the full-text content. CENTRAL Central Register of Controlled Trials.

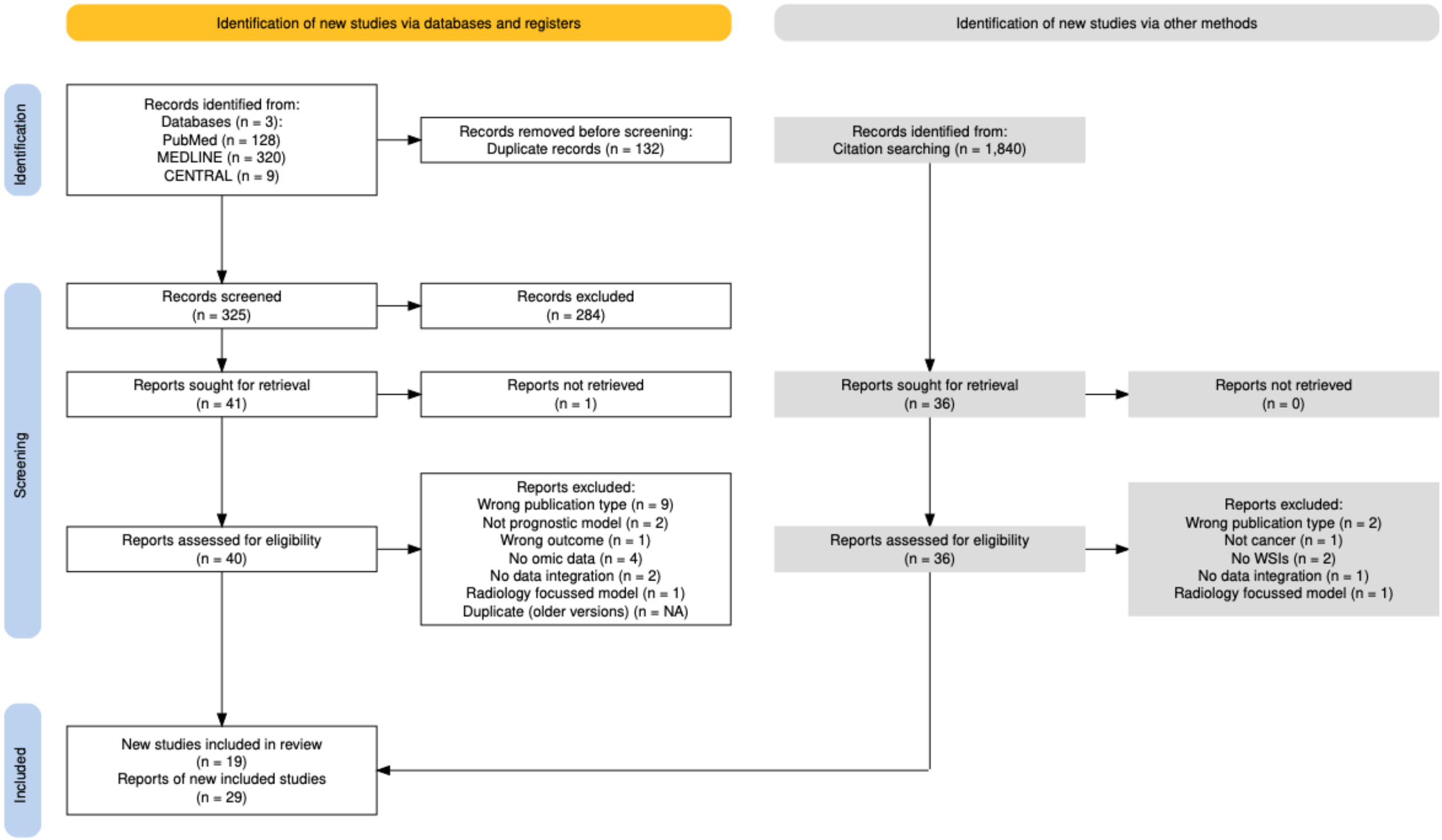

**Figure 2. PROBAST+AI summary results.** Summarised results for the assessments of (a) quality and (b) applicability (study development), and (c) risk of bias and (d) applicability (study evaluation) for the 48 studies in this review.

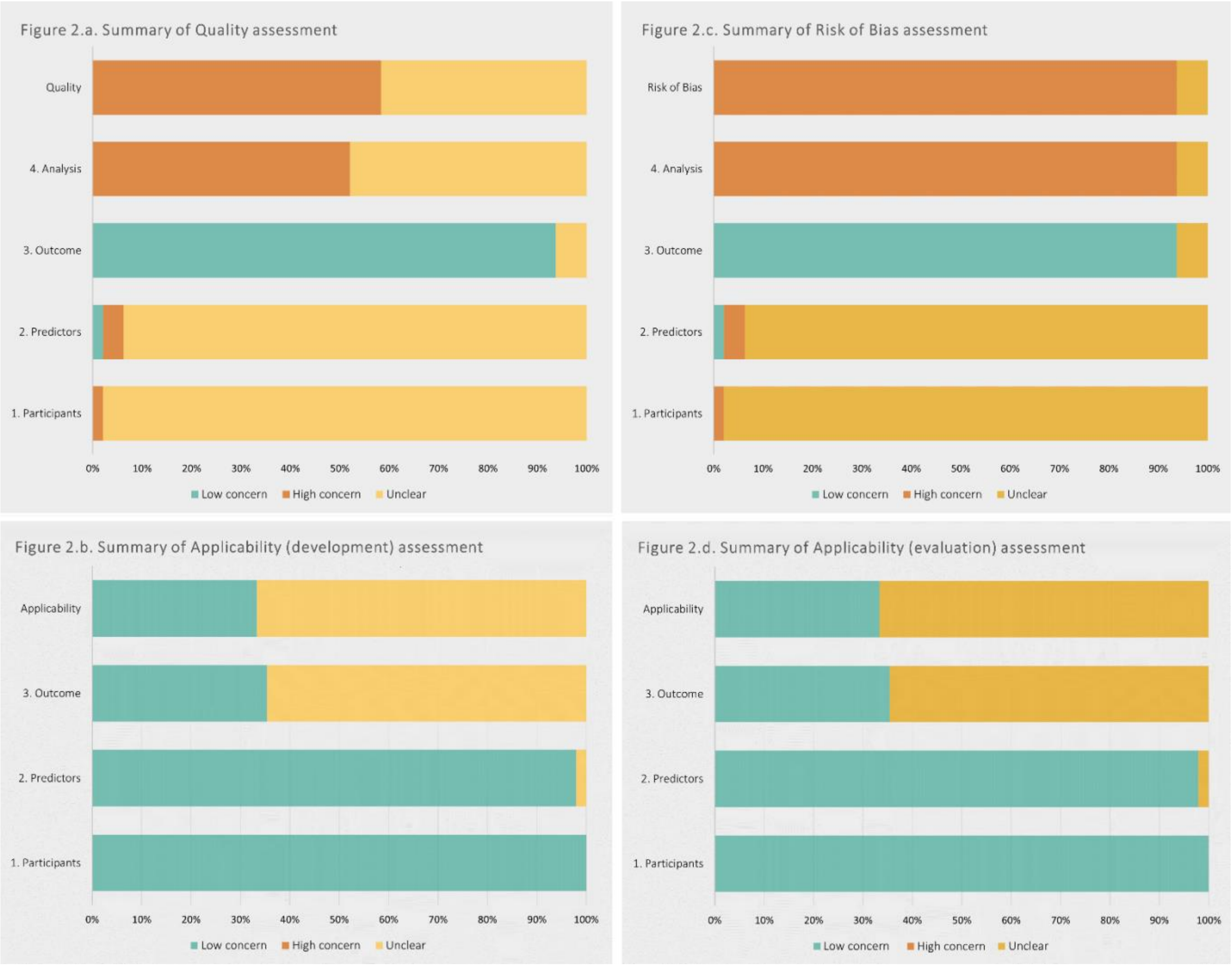

**Figure 3. Cancer types represented across the studies in this review.**

The number of studies (y-axis) are presented by organ systems (x-axis), showing a differentiation between whether they were part of multi- or single cancer studies. Nineteen studies developed models across multiple cancer subtypes and organs, with the remaining 29 being single cancer type studies. The most studied cancers were those of the brain, breast and lung.

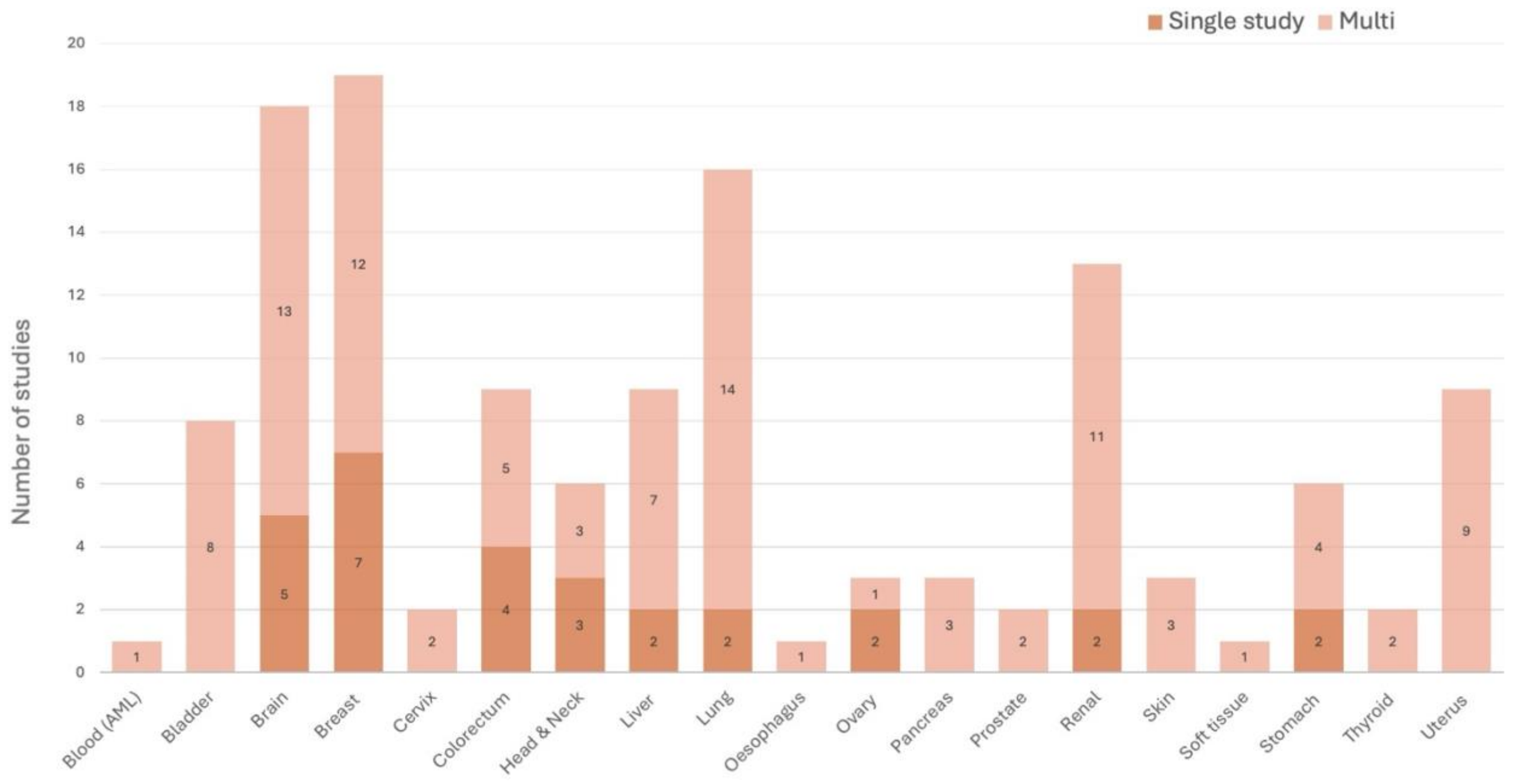

**Figure 4. Comparison of multimodal and unimodal model performance using c-index**. The 25 studies which reported the c-index for unimodal comparison are shown above. Studies are ordered by year of publication (x-axis) with regression lines to show a trend of performance (c-index, y-axis) over time. Different studies performed different unimodal comparisons reflected by incomplete data points for several of the studies. Where multiple omics were evaluated separately, the best performing is charted here. A further 7 studies evaluated unimodal performance against the multimodal model but generated AUC (n=6) or F1 metrics (n=1) so were not included in this comparison. Table S1. indicates all studies which undertook unimodal performance comparisons.

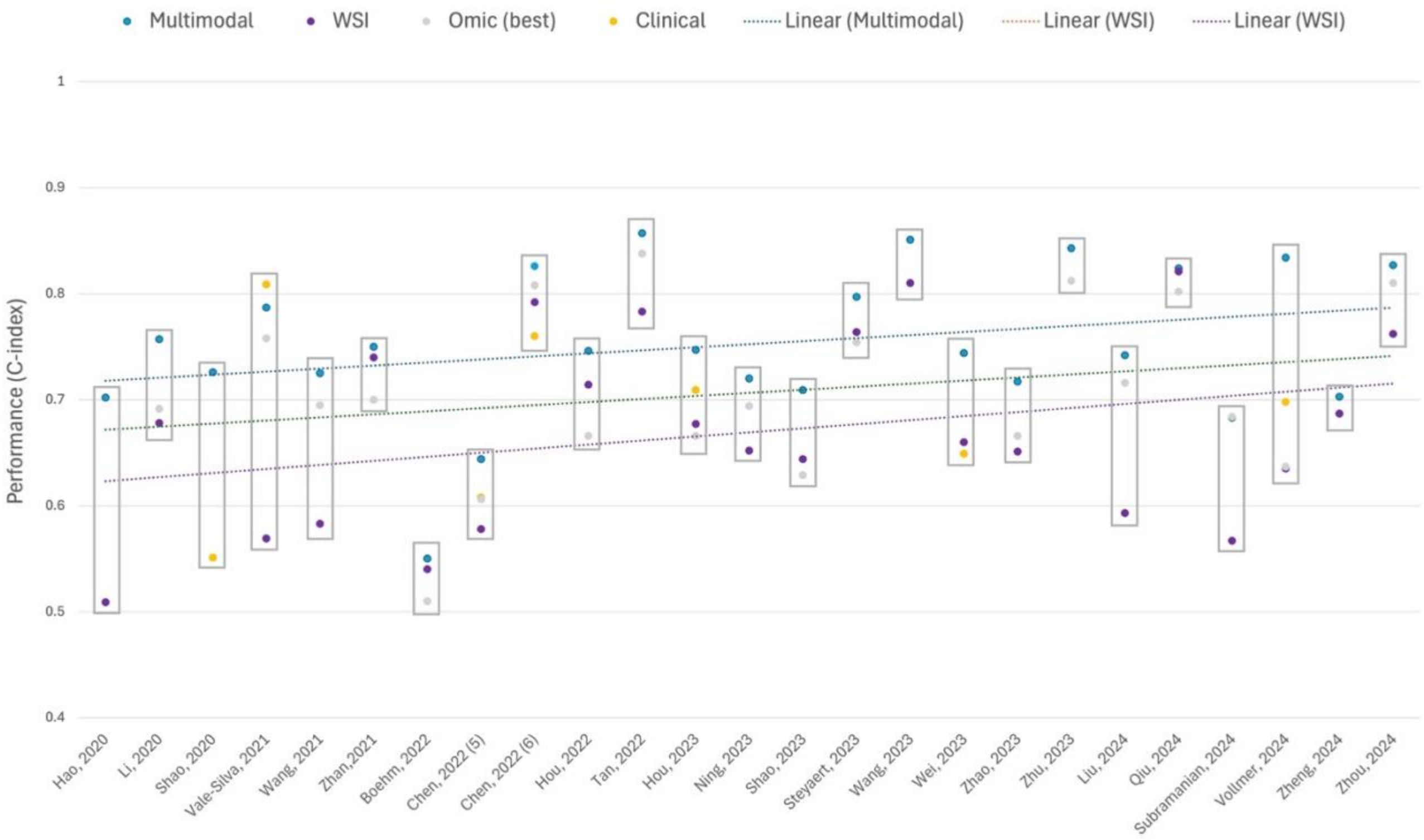

## Table 1: Characteristics of the studies included in the systematic review.

| Author, Year | Location | Model of interest | Cancer types | Internal data | | Ext. Data | WSI data | Omic data | Clinical data | Participant characteristics provided in study: | | | Data availability | Code available |
|---|---|---|---|---|---|---|---|---|---|---|---|---|---|---|
| | | | | Name | Origin | Name (origin) | Fixation Stain Resolution | Data types | Variables in model | Number of participants | Demo-graphic information | Disease information | | |
| *Boehm, 2022 (22)* | *USA* | *GHC** | *(Ovary) HGSOC* | *MSKCC; TCGA-OV* | *USA, USA** | *-* | *FFPE H&E -* | *CNV, somatic mutation, SNP* | *RD status, PARP inhibitor administration* | *283 (444)* | *Age, gender, race* | *Stage* | *Partially* | *Yes* |
| *Cheerla, 2019 (23)* | *USA* | *Multimodal (Clin+ miRNA+ mRNA+ WSI)* | *(Multi - 20) Pancancer* | *TCGA-PanCanAtlas* | *USA** | *-* | *- H&E -* | *mRNA, miRNA* | *Race, age, gender, grade* | *11160* | *-* | *-* | *All open source** | *Yes* |
| *Chen, 2021 (24)* | *China* | *HTRS* | *(Head & Neck) SCC* | *TCGA* | *-* | *** | *- H&E 20x, 40x* | *mRNA* | *-* | *212* | *Age, gender* | *Tumour site, grade, stage* | *On request* | *-* |
| *Chen, 2021 (25)* | *China* | *Multiomics model* | *(Lung) Adenocarcinoma* | *TCGA (LUAD)* | *USA** | *** | *FFPE H&E 40x* | *mRNA, genomics, protein expression* | *-* | *470* | *Age, gender* | *Stage* | *All open source* | *-* |
| *Chen, 2022 (26)* | *USA* | *MMF* | *(Multi - 14) Pancancer* | *TCGA (BLCA, BRCA, COADREAD, HNSC, KIRC, KIRP, LGG, LIHC, LUAD, LUSC, PAAD, SKCM, STAD, UCEC)* | *USA** | *-* | *FFPE H&E -* | *CNV, somatic mutation, mRNA* | *-* | *5720* | *Age, gender, ethnicity* | *Grade* | *All open source* | *Yes* |
| *Chen, 2022 (27)* | *USA* | *Pathomic Fusion* | *(Multi - 2) Glioma* | *TCGA (LGG, GBM, KIRC)* | *USA** | *-* | *FFPE* H&E -* | *CNV, somatic mutation, mRNA* | *-* | *769 (1186)* | *-* | *-* | *All open source** | *Yes* |
| *Cheng, 2017 (28)* | *China* | *Lasso-Cox* | *Renal (CCRCC)* | *TCGA* | *USA** | *-* | *- H&E 20x, 40x* | *mRNA* | *-* | *410* | *Age, gender* | *Grade, Stage* | *Yes* | *Yes** |
| *Hao, 2020 (29)* | *USA* | *PAGE-NET* | *(Brain) Glioblastoma* | *TCGA-GBM* | *USA** | *-* | *Frozen H&E* 20x* | *mRNA* | *Age* | *447* | *Age* | *-* | *All open source** | *Yes* |
| *Hou, 2022 (30)* | *China* | *Multi-modality* | *(Liver) HCC* | *TCGA* | *USA** | *-* | *FFPE H&E 5-40x* | *mRNA* | *Age, Sex* | *346* | *Age, Gender* | *T group, N group, M group and overall TNM stage* | *All open source* | *Yes* |
| *Hou, 2023 (31)* | *China* | *HGCN* | *(Multi - 6) KIRC* | *TCGA* | *USA** | *-* | *FFPE* H&E* -* | *"Genomic profile data"* | *Age, Gender, Therapy, BMI* | *385(2102)* | *-* | *-* | *All open source** | *Yes* |
| Ji, 2024 (32) | China | HGRS | (Renal ) CCRCC | TCGA | USA* | - | FFPE H&E 20x, 40x | mRNA | - | 281 | Age, gender | TNM stage, Grade | All open source | - |
| Li, 2020 (33) | China | DeepHit | (Breast) - | TCGA-BRCA | USA* | - | - H&E* - | mRNA | - | 826 | - | - | All open source | - |
| Li, 2021 (34) | China | HGPF | (Colon) Adenocarcino ma | TCGA-COAD | USA* | - | FFPE H&E* - | mRNA | - | 199 | Age, gender | Stage/TNM classification | All open source | - |
| Li, 2022 (35) | China | HFBSurv | (Multi -10) Breast | TCGA | USA* | - | - H&E 40x | mRNA, CNV | - | 1015 (-) | - | - | All open source* | Yes |
| Liu, 2023 (36) | China | MGCT | (Multi - 5) Pancancer | TCGA (LUAD, BRCA, BLCA, GBMLGG, UCEC) | USA* | - | - H&E* - | CNV, somatic mutation, mRNA | - | 3557 | - | - | All open source* | Yes |
| Liu, 2024 (37) | China | IntraSA-InterCA | (Breast) - | TCGA | USA* | - | - H&E* 40x | mRNA | - | 345 | - | - | All open source* | - |
| Lv, 2021 (38) | China | PG-TFNet | (Colorectum) - | TCGA | USA* | - | - H&E - | mRNA | Age, gender, pathologic stage | 522 | - | - | All open source* | - |
| Lv, 2023 (39) | China | Trans-Surv | (Colorectum) - | TCGA; (NCT-CRC-HE-100K; CRC-VAL-HE-7K tissue classifier used in model) | USA* | - | - H&E - | CNV, mRNA | Age, gender, pathologic stage | 520 | - | - | All open source* | - |
| Ning, 2020 (40) | China | Gene+His* | (Renal) CCRCC | TCGA | USA* | - | - H&E* - | mRNA | - | 209 | Age, gender | Grade, stage | All open source* | Yes |
| Ning, 2023 (41) | China | McLR Framework | (Multi - 3 ) LIHC | TCGA (KIRC, LIHC, LUAD) | USA* | - | - H&E - | mRNA | - | 334 (1177) | Age, gender | Stage, pT stage | All open source* | Yes* |
| Perez-Herrera, 2024 (42) | Spain | Model 3 | (Breast) - | TCGA (BRCA) & TIGER; (segmentation training only); METABRIC | USA*; Netherland s & Belgium | METABRI C (Canada-UK*) | - H&E - | SNP, CNV, mRNA | Age, subtype | 3593 | Age | Subtype | All open source* | - |
| Qiu, 2024 (43) | China | DDM-net | (Brain) Glioma | TCGA (GBMLGG) | USA* | - | FFPE H&E - | CNV, somatic mutation, mRNA | - | 954 | - | WHO grading | All open source | - |
| Shao, 2020 (44) | China | OMMFS | (Multi - 3) LUSC | TCGA (KIRC, KIRP, LUSC) | USA* | - | - H&E* - | CNV, DNAm, mRNA | - | 365 (787) | - | Stage | All open source* | - |
| Shao, 2023 (45) | China | FAM3L | (Multi - 3) Multi (KIRP) | TCGA (BRCA, KIRP, KIRC) | USA* | - | - H&E* - | mRNA | - | 251 (1824) | Age | - | All open source* | - |
| Shao, 2020 (46) | China | M2DP | (Multi - 3) BRCA | TCGA (LUSC, BRCA, LIHC) | USA* | - | - H&E* - | mRNA | Stage | 790 (1324) | - | Stage | All open source* | - |
| Shao, 2023 (47) | China | IMO-TILS | (Multi - 3) Breast | TCGA | USA* | - | -<br>H&E*<br>- | mRNA, miRNA | - | 365 (920) | Age | - | All open source* | - |
| Steyaert, 2023 (48) | USA | Late fusion | (Brain - Multi) Adult Glioblastoma | TCGA; Paediatric Brain Tumour Atlas (PBTA) Kids First cohort | USA* | CPTAC-GBM (USA*) | FFPE<br>H&E*<br>- | mRNA | - | 783 (1088) | - | Low grade, high grade | All open source | Yes |
| Subramanian, 2021 (49) | USA | GCCA | (Breast) - | TCGA (BRCA) | USA* | - | -<br>H&E*<br>- | mRNA | - | 974 | - | - | All open source | Yes |
| Subramanian, 2024 (50) | USA | SCCA | (Breast) - | TCGA (BRCA) | USA* | - | -<br>H&E<br>- | mRNA | - | 974 | Age, gender, ethnicity | - | Yes | Yes |
| Sun, 2018 (51) | China | GPMKL | (Breast) - | TCGA (BRCA) | USA* | - | -<br>H&E<br>40x | CNV, DNAm, mRNA protein expression | - | 578 | Age | - | All open source* | - |
| Tan, 2022 (52) | China | MultiCoFusion | (Brain) GBMLGG | TCGA (GBM, LGG) | USA* | - | -<br>H&E<br>- | mRNA | Grade | 469 | - | Grade | All open source* | - |
| Vale-Silva, 2021 (53) | Germany | MultiSurv | (Multi - 33) Multi | TCGA | USA* | - | -<br>H&E<br>- | CNV, DNAm, mRNA, miRNA, DNAm | Age, gender, race, cancer type, tumour stage, prior malignancy, synchronous malignancy, prior treatment, pharmaceutical treatment, radiation treatment. | 11167 | - | - | All open source* | Yes |
| Vollmer, 2024 (54) | Germany | Random Survival Forest | (Oral) SCC | TCGA (HNSC) | USA* | - | -<br>H&E<br>40x | mRNA, somatic/germline DNA | Yes but unclear what | 406 | Age, gender, race, ethnicity | Stage | All open source | Yes |
| Wang, 2021 (55) | China | GPDBN | (Breast) - | TCGA (BRCA) | USA* | - | -<br>H&E*<br>40x | mRNA | - | 1015 | - | - | All open source* | Yes |
| Wang, 2023 (56) | China | HC-MAE | (Multi - 6 ) LGG | TCGA (LIHC, BRCA, LUAD, COAD, LGG, STAD) | USA* | - | -<br>H&E*<br>- | DNAm, mRNA, miRNA | - | 451 (2420) | - | - | All open source* | Yes |
| Wei, 2023 (57) | China | MultiDeepCox-SC | (Stomach) - | TCGA (STAD) | USA* | * | FFPE<br>H&E<br>20x | mRNA | Age | 357 | Age, gender, race | Stage, grade | All open source | - |
| Wu, 2023 (58) | China | CAMR | (Multi - 3) LGG | TCGA (BRCA, LUSC, LGG) | USA* | - | -<br>H&E*<br>40x | CNV, mRNA | - | 629 (2135) | - | - | All open source* | Yes |
| Xie, 2024 (59) | China | GaCaMML | (Stomach) | Ruijin Hospital (Shanghai); GSE54129 (Public Chinese cohort, Ruijin) | China, China | TCGA-LUAD - (USA*) | -<br>H&E<br>40x | mRNA | - | 63 (95) | Age, sex | Histological type, grade, TNM stage, Bormann classification, tumour position | On request | - |
| Zeng, 2020 (60) | China | Multi-omics model | (Head & Neck) SCC | TCGA (HNSCC) | USA* | - | -<br>H&E<br>20x, 40x | mRNA, somatic mutation, protein expression | - | 216 | Age, gender | Anatomic site, stage | All open source | - |
| Zeng, 2021 (61) | China | Multi-omics model | (Ovary) HGSOC | TCGA | USA* | * | -<br>H&E<br>- | Genomics, mRNA, protein expression | - | 229 | Age | Stage | All open source* | - |
| Zhan, 2021 (62) | China | Two-stage Cox-nnet | (Liver) HCC | TCGA | USA* | - | FFPE<br>H&E<br>- | mRNA | - | 290 | - | - | All open source* | Yes |
| Zhang, 2020 (63) | China | HI-MKL | (Brain) Glioblastoma | TCGA (GBM) | USA* | - | -<br>H&E*<br>20x, 40x | CNV, DNAm, mRNA | - | 251 | Age, gender | - | All open source* | - |
| Zhao, 2023 (64) | China | Ada-RSIS | (Multi - 3 ) UCEC | TCGA (BLCA, UCEC, LGG) | USA* | TCGA (GBM, UCS) (USA*) | -<br>H&E<br>- | mRNA | - | 539 (1437) | Age, gender | Stage | All open source* | Yes* |
| Zheng, 2024 (65) | USA | FSM | (Lung - Multi) LUAD | TCGA; National Lung Screening Trial (generating features) | USA* | CPTAC (LUAD, LUSC) (USA*) | -<br>H&E*<br>- | mRNA | - | 444 (1259) | Age, gender, race | Stage | All open source* | Yes |
| Zhou, 2023 (66) | Hong Kong | CMTA | (Multi - 5) GBMLGG | TCGA (BLCA, BRCA, GBMLGG, LUAD, UCEC) | USA* | - | -<br>H&E<br>40x | CNV, mRNA, SNP | - | 569 (2831) | - | - | All open source* | Yes |
| Zhou, 2023 (67) | USA | Integrative CNN | (Colon) Adenocarcinoma | TCGA (COAD); Wayne State University | USA*, USA | TCGA-READ (USA*) | FFPE<br>H&E<br>- | Somatic mutation, MSI signature | Age, gender, TNM stage, T stage, N, stage, M stage | Unclear | Age, sex | Stage | Yes | Yes |
| *Zhou, 2024 (68)* | *Hong Kong* | *MSEN* | *(Multi - 5 ) GBM-LGG* | *TCGA (BLCA, BRCA, LUAD, UCEC, GBMLGG)* | *USA** | *-* | *- H&E* -* | *CNV, Somatic mutation, mRNA* | *-* | *492 (2670)* | *-* | *-* | *All open source** | *Yes** |
| *Zhu, 2023 (69)* | *China* | *SAMMS* | *(Multi - 2) LGG* | *TCGA (LGG, KIRC)* | *USA** | *-* | *- H&E* 20x* | *CNV, mRNA, miRNA* | *Age, Gender* | *500 (806)* | *-* | *-* | *All open source** | *-* |

**Asterisk** generally used to denote fields where answer reasonably assumed by both investigators. - used where information not provided or unclear.
***Model of interest (asterisk)*** - indicates where an ineligible multimodal model (e.g. incorporating radiology) achieved a higher result than our model of interest.
***External data (asterisk)*** where external validation datasets are used within the study but not for the multimodal model.
***Code availability (asterisk)*** – where code links could not find associated code despite described as available
**Abbreviations**: BMI (body mass index), CNV (Copy number variation), CPTAC (Clinical Proteomic Tumour Analysis Consortium), DNAm (DNA methylation), FFPE (Formalin fixed paraffin embedded), H&E (Haematoxylin and Eosin), mRNA (messenger RNA), miRNA (micro RNA), MSKCC (Memorial Sloan Kettering Cancer Centre), SNP (Single nucleotide polymorphisms), TCGA (The Cancer Genome Atlas).

## Table 2: Characteristics of the models of interest included in the systematic review.

| Method | Author, Year (ref) | Modelling description | Study size | Events | Predictors | | WSI Feature | Omic feature | Fusion level | Validation | Performance measures | | | | Internal results | | | |
|---|---|---|---|---|---|---|---|---|---|---|---|---|---|---|---|---|---|---|
| | | | n | n (%) | Cand. | Final | Approach (Named method) | Named method (s) | | | Calibration | Discrimination | Overall | Clin. Utility | C-index | Variability (metric) | AUC | Variability (metric) |
| *ML regularised regression* | *Cheng, 2017 (28)* | ***Lasso-Cox*** *Image features and eigengenes into lasso-regularised Cox Proportional Hazards model* | *410* | *135 (33)* | *165* | *13* | *Hand crafted -* | *lmQCM* | *Feature* | *Cross-validation* | *-* | *Log-rank test / Risk group curves* | *-* | *-* | *-* | *-* | *-* | *-* |
| | *Shao, 2020 (44)* | ***OMMFS*** *Feature selection via ordinal multi-modality feature selection model, then features into Cox Proportional Hazards model* | *365* | *145 (40)* | *2011* | *16* | *Hand crafted -* | *WGCNA* | *Feature* | *Random split data* | *-* | *C-Statistic / AUC graph / Log-rank test / Risk group curves* | *-* | *-* | *0.773* | *x* | *0.793* | *x* |
| | *Shao, 2020 (46)* | ***M2DP*** *Multi-task, multi-modal feature selection algorithm with Cox Proportional Hazards model* | *790* | *82 (10)* | *-* | *155* | *Hand crafted -* | *lmQCM* | *Feature* | *Cross-validation* | *-* | *C-Statistic / Log-rank test / Risk group curves* | *Brier score* | *-* | *0.726* | *±0.06 (unclear)* | *-* | *-* |
| | *Shao, 2023 (45)* | ***FAM3L*** *Feature-aware multi-modal metric learning method learns survival associated features which are combined into Cox Proportional Hazards model* | *251* | *38 (15)* | *-* | *163* | *Hand crafted -* | *lmQCM* | *Feature* | *Cross-validation* | *-* | *C-Statistic / AUC graph / Log-rank test / Risk group curves* | *-* | *-* | *0.752* | *±0.061 (unclear)* | *0.791* | *0.069 (unclear)* |
| *Machine learning* | *Sun, 2018 (51)* | ***GPMKL** Selected features integrated by multiple kernel learning, with a separate kernel per data modality* | *578* | *445 (77)* | *59428* | *350* | *Hand crafted CellProfiler* | *-* | *Feature* | *Cross-validation* | *-* | *C-Statistic / AUC graph* | *-* | *-* | *0.643* | *x* | *0.828* | *±0.034 (St Err)* |
| | *Zeng, 2020 (60)* | ***Multi-omics model** Extracted image and omic features integrated directly into random forest model* | *216* | *106 (49)* | *35302* | *954* | *Hand crafted CellProfiler* | *-* | *Feature* | *Cross-validation* | *-* | *AUC graph / Log-rank test / Risk group curves* | *-* | *DCA* | *-* | *-* | *0.929 (5-yr)* | *x* |
| | *Zhang, 2020 (63)* | ***HI-MKL** Histopathological image feature integrating multiple kernel learning method - multiple kernels derived from selected features are combined within each data type, before combining across modalities for a combined kernel for MKL.* | *251* | *204 (81)* | *66650* | *350* | *Hand crafted CellProfiler* | *-* | *Feature* | *Cross-validation* | *-* | *AUC graph / Risk group curves* | *-* | *-* | *-* | *-* | *0.932* | *x* |
| | *Chen, 2021 (25)* | ***Multiomics model** Selected histopathology image and multiomics features (multiple selection techniques) integrated by Random Forest model* | *470* | *169 (36)* | *-* | *954* | *Hand crafted CellProfiler* | *-* | *Feature* | *Cross-validation* | *-* | *AUC graph / Log-rank test / Risk group curves* | *-* | *DCA* | *-* | *-* | *0.938 (3-yr)* | *x* |
| | *Chen, 2021 (24)* | ***HTRS** Selected histopathology image and gene features integrated by a Random Forest model* | *212* | *-* | *5532* | *259* | *Hand crafted CellProfiler* | *WGCNA* | *Feature* | *Cross-validation* | *Calib. plot* | *C-Statistic / AUC graph / Log-rank test / Risk group curves* | *-* | *DCA* | *0.768* | *0.715-0.820 (95% CI)* | *0.826 (5-yr)* | *x* |
| | *Li, 2021 (34)* | ***HGPF** Direct integration of features selected by SVM-RFE and LASSO-COX (images), and WGCNA (hub genes) into Random Forest* | *199* | *32 (16)* | *20344* | *377* | *Hand crafted CellProfiler* | *WGCNA* | *Feature* | *Cross-validation* | *Calib. plot* | *AUC graph / Log-rank test / Risk group curves* | *-* | *DCA* | *-* | *-* | *0.924 (5-yr)* | *x* |
| | *Subramanian, 2021 (49)* | ***GCCA** Graph-structured variant of sparse canonical correlation analysis, capturing intra and intermodality correlations* | *974* | *-* | *-* | *3213* | *Hand crafted CellProfiler* | *-* | *Feature* | *Cross-validation* | *-* | *-* | *F1 score* | *-* | *-* | *-* | *-* | *-* |
| | *Zeng, 2021 (61)* | ***Multi-omics model** Extracted image features and selected omics features integrated directly into Random Forest model* | *229* | *133 (58)* | *-* | *970* | *Hand crafted CellProfiler* | *-* | *Feature* | *Cross-validation* | *-* | *AUC graph / Log-rank test / Risk group curves* | *-* | *DCA* | *-* | *-* | *0.911 (5-yr)* | *x* |
| | Ning, 2023 (41) | ***McLR Framework*** *Unimodal features concatenated into multimodal feature matrix and used as input to multi-constraint latent representation framework* | *334* | *116 (35)* | *20003* | *256* | *Hand crafted PFTAS* | *-* | *Feature* | *Cross-validation* | *-* | *C-Statistic / Log-rank test / Risk group curves* | *-* | *-* | *0.72* | *0.045 (StD)* | *-* | *-* |
| | *Zhao, 2023 (64)* | ***Ada-RSIS*** *Adaptive risk-aware sharable and individual subspace learning - subspace learning of modality specific and modality sharable representations passed through function to capture intra and inter modality terms. Representations integrated via grouping co-expression constraint and gaussian-based weighting strategy.* | *539* | *91 (17)* | *20003* | *256* | *Hand crafted PFTAS* | *-* | *Feature* | *Cross-validation* | *-* | *C-Statistic / Risk group curves* | *-* | *-* | *0.711* | *±0.048 (StD)* | *-* | *-* |
| | *Ji, 2024 (32)* | ***HGRS*** *Direct integration of features selected by SVM-RFE and LASSO-COX (images), and WGCNA (hub genes) into Random Forest survival model* | *281* | *76 (27)* | *-* | *8* | *Hand crafted CellProfiler* | *WGCNA* | *Hybrid* | *Cross-validation* | *Calib. plot* | *AUC graph / Log-rank test / Risk group curves* | *-* | *DCA* | *-* | *-* | *0.780 (5-yr)* | *x* |
| | *Subramanian, 2024 (50)* | ***SCCA*** *Probabilistic graph modelling by penalised canonical correlation analysis-based joint embeddings and Cox Proportional Hazards model* | *974* | *-* | *21075* | *2075* | *Hand crafted CellProfiler* | *-* | *Feature* | *Cross-validation* | *-* | *C-Statistic / Risk group curves* | *-* | *-* | *0.683* | *±0.0669 (unclear)* | *-* | *-* |
| | *Vollmer, 2024 (54)* | ***Random Survival Forest*** *Feature selection via then Random Forest model for survival prediction.* | *406* | *177 (44)* | *-* | *242* | *Hand crafted CellProfiler* | *-* | *Feature* | *Cross-validation* | *-* | *C-Statistic* | *-* | *-* | *0.834* | *x* | *-* | *-* |
| *Deep learning* | *Cheerla, 2019 (23)* | ***Multimodal (Clin + miRNA + mRNA + WSI)*** *Modality specific features generated via multiple CNNs compressed by unsupervised encoder into single vector for survival prediction by cox loss function. Multimodal dropout* | *11160* | *-* | *62308* | *2048* | *Learned SqueezeNet CNN* | *Deep Highway Neural Network* | *Feature* | *Random split data* | *-* | *C-Statistic* | *-* | *-* | *0.78* | *x (average result)* | *-* | *-* |

| | | | | | | | | | | | | | | | | | |
|---|---|---|---|---|---|---|---|---|---|---|---|---|---|---|---|---|---|
| | approach used in training. | | | | | | | | | | | | | | | | |
| Hao, 2020 (29) | ***PAGE-NET*** *Pathology (CNN), genome and clinical (adapted CoxPASNet) layers fused in Cox Proportional Hazards model* | 447 | - | 6405 | 81 | Learned Pre-trained CNN | CoxPASNET | Feature | Cross-validation | - | C-Statistic / Wilcoxon rank-sum test | - | - | 0.702 | ±0.0294 (unclear) | - | - |
| Li, 2020 (33) | ***DeepHit*** *Fully connected neural network with image and omic feature fusion via conditional autoencoder.* | 826 | - | - | - | Learned ResNet, DCGMM | - | Feature | Cross-validation | - | C-Statistic | - | - | 0.757 | 0.713-0.801 (95% CI) | - | - |
| Ning, 2020 (40) | ***Gene+His*** *Cross-modal feature-based integrative framework. Deep features extracted from images (CNNs), combined with eigengenes (WGCNA) and fed into cox model for prediction* | 209 | 60 (29) | - | 138 | Learned CNN | WGCNA | Feature | Cross-validation | - | C-Statistic / Log-rank test / Risk group curves | - | - | 0.638 | 0.527-0.749 (unclear) | - | - |
| Lv, 2021 (38) | ***PG-TFNet*** *Output of transformer-based multiscale pathological feature fusion module (images) and multilayer perceptron features (genes) integrated by cross attention transformer-based multimodal feature fusion module and fed into cox layer (with clinical data) for survival prediction.* | 522 | - | - | - | Learned ResNet (multiscale) | MLP | Feature | Cross-validation | - | C-Statistic / Log-rank test / Risk group curves | - | - | 0.816 | ±0.016 (unclear) | - | - |
| Vale-Silva, 2021 (53) | ***MultiSurv*** *Dedicated deep learning submodels per data modality (n=6), fed into data fusion layer then final network takes fused feature representation and outputs conditional survival probabilities.* | 11081 | 3601 (32) | 108437 | 3072 | Learned ResNet50 CNN (ImageNet) | CNN | Feature | Bootstrap | - | C-Statistic / Log-rank test / Risk group curves | Brier score | - | 0.787 | 0.769-0.806 (95% CI) | - | - |

| | | | | | | | | | | | | | | | | | | |
|---|---|---|---|---|---|---|---|---|---|---|---|---|---|---|---|---|---|---|
| | *Wang, 2021 (55)* | ***GPDBM** Outputs of modality specific bilinear feature encoding modules (BFEM) and an inter-modality BFEM fused and fed into fully connected neural network for survival prediction* | *1015* | *-* | *20446* | *64* | *Hand crafted CellProfiler* | *-* | *Feature* | *Cross-validation* | *-* | *C-Statistic / AUC graph / Log-rank test / Risk group curves* | *Sensitivity, specificity, accuracy, precision, F1* | *-* | *0.725* | *±0.066 (unclear)* | *0.808* | *±0.053 (unclear)* |
| | *Zhan, 2021 (62)* | ***Two-stage Cox-nnet** Outputs of modality specific cox-nnet models integrated as nodes to a second stage cox-nnet with an output layer for cox regression.* | *290* | *-* | *-* | *-* | *Hand crafted CellProfiler* | *-* | *Feature* | *Cross-validation* | *-* | *C-Statistic / Log-rank test / Risk group curves* | *-* | *-* | *0.75* | *x* | *-* | *-* |
| | *Boehm, 2022 (22)* | ***GHC** Fusion of negative log partial hazards from unimodal cox models with learned image features into a multimodal linear cox proportional hazards model* | *283* | *-* | *-* | *5* | *Combo ResNet18* | *HRD status inferred* | *Decision* | *Random split data* | *-* | *C-Statistic / Log-rank test / Risk group curves / Kendall rank correlation* | *-* | *-* | *0.55* | *0.531-0.563 (95% CI)* | *-* | *-* |
| | *Chen, 2022 (26)* | ***MMF** Kronecker product fusion of unimodality subnetworks to generate multimodal representation feature and predict survival via log likelihood function* | *5720* | *1651 (29)* | *-* | *64* | *Learned ResNet50 (ImageNet), attention module* | *SNN* | *Feature* | *Cross-validation* | *-* | *C-Statistic / AUC graph / Log-rank test / Risk group curves* | *-* | *-* | *0.644* | *x (average result)* | *0.662* | *x (average result)* |
| | *Chen, 2022 (27)* | ***Pathomic Fusion** Unimodal networks as feature extractors (CNN and GCN - images, Feed-forward SNN - genomic) fused by gating attention mechanism and Kronecker product to model modality interactions. Network supervised by cox partial likelihood loss function.* | *769* | *-* | *-* | *-* | *Combo VGG19 (ImageNet), GCN* | *SNN* | *Feature* | *Cross-validation* | *-* | *C-Statistic / Log-rank test / Risk group curves* | *-* | *-* | *0.826* | *±0.009 (unclear)* | *-* | *-* |
| | *Hou, 2022 (30)* | ***Multi-modality** Pathology patch level risk score (via multi-instance fully convolutional network aggregated to patient level by attention mechanism) integrated with hub genes into Cox* | *346* | *123 (36)* | *5005* | *8* | *Learned VGG19* | *WGCNA* | *Hybrid* | *Cross-validation* | *Calibration plot* | *C-Statistic / AUC graph / Log-rank test / Risk group curves* | *-* | *DCA* | *0.746* | *±0.077 (95% CI)* | *0.816 (1-yr)* | *x* |
| | | *Proportional Hazards model.* | | | | | | | | | | | | | | | | |
| | *Li, 2022 (35)* | ***HFBSurv*** *Fusion of modality-specific and cross-modality attentional factorized bilinear modules with cox partial likelihood loss function for survival prediction* | *1015* | *-* | *46125* | *240* | *Hand crafted CellProfiler* | *-* | *Feature* | *Cross-validation* | *-* | *C-Statistic / AUC graph / Log-rank test / Risk group curves* | *-* | *-* | *0.766* | *±0.024 (unclear)* | *0.806* | *±0.025 (unclear)* |
| | *Tan, 2022 (52)* | ***MultiCoFusion*** *Features concatenated then fused via three-layer FCNN before output fed to multitask output.* | *469* | *-* | *62437* | *2000* | *Learned ResNet-152 (ImageNet)* | *SGCN* | *Feature* | *Cross-validation* | *-* | *C-Statistic / AUC graph / Risk group curves* | *-* | *-* | *0.857* | *±0.015 (unclear)* | *0.915* | *±0.016 (unclear)* |
| | *Hou, 2023 (31)* | ***HGCN*** *Distinct modality graph representations fed into Graph Convolutional Network layers then hypergraph convolutional layer with hyperedge mixing module (intermodal interaction) before decision via cox loss. Jointly trained with online masked autoencoder (to simulate missing data)* | *385* | *-* | *21504* | *3* | *Learned KimiaNet (CNN)* | *GSEA* | *Decisio n* | *Cross-validation* | *-* | *C-Statistic / Log-rank test / Risk group curves* | *Brier score* | *-* | *0.747* | *±0.007 (95% CI)* | *-* | *-* |
| | *Liu, 2023 (36)* | ***MGCT*** *Modality specific embeddings integrated with two-stage mutual guided cross-modality transformer framework* | *3557* | *-* | *-* | *256* | *Learned ResNet50 (ImageNet)* | *SNN* | *Feature* | *Cross-validation* | *-* | *C-Statistic / AUC graph / Log-rank test / Risk group curves* | *-* | *-* | *0.663* | *x (average result)* | *-* | *-* |
| | *Lv, 2023 (39)* | ***Trans-Surv*** *Multi Scale Fusion Transformer (image feature extraction), integration of image and omic data by cross-modal fusion transformer, combined with clinical data into Cox Layer* | *520* | *-* | *-* | *-* | *Learned Multiple CNNs* | *-* | *Feature* | *Cross-validation* | *-* | *C-Statistic / Log-rank test / Risk group curves* | *-* | *-* | *0.822* | *±0.023 (unclear)* | *-* | *-* |
| | Shao, 2023 (47) | ***IMO-TILS*** *Deep generalised canonical correlation analysis with attention layer to fuse image and multi-omic data* | 365 | 32 (9) | 272 | 272 | Learned Unet++, ResNet-101 (ImageNet), KNN | lmQCM | Feature | Cross-validation | - | C-Statistic / AUC graph / Log-rank test / Risk group curves | - | - | 0.709 | ±0.030 (unclear) | 0.737 | ±0.047 (unclear) |
| | Steyaert, 2023 (48) | ***Late Fusion*** *Modality specific cox proportional hazards models using model derived image (CNN) and omic (MLP) features. Outputs integrated into a final cox regression module (late fusion)* | 783 | - | 12879 | 4096 | Learned ResNet50 CNN | MLP | Decision | Cross-validation | - | C-Statistic / Log-rank test / Risk group curves | Brier score, Composite score | - | 0.797 | ±0.019 (StD) | - | - |
| | Wang, 2023 (56) | ***HC-MAE*** *Hierarchical cross-attention masked autoencoder - multiscale WSI representations (guided by omic cross attention)and omic features fed through separate transformers then fused via concatenation. Survival function negative log partial likelihood.* | 451 | - | - | 2052 | Learned Vision Transformer-based masked Autoencoder | MLP | Feature | Cross-validation | - | C-Statistic / Log-rank test / Risk group curves | - | - | 0.851 | ±0.033 (unclear) | - | - |
| | Wei, 2023 (57) | ***MultiDeepCox-SC*** *Integration of age and gene predictors with risk score of deep learning-based image model (DeepCoxSC) using a cox model* | 357 | 142 (40) | 61584 | 12 | Hand crafted CellProfiler | - | Hybrid | Cross-validation | - | C-Statistic / AUC graph / Log-rank test / Risk group curves | - | - | 0.744 | ±0.070 (unclear) | 0.833 (2-yr) | ±0.055 (unclear) |
| | Wu, 2023 (58) | ***CAMR*** *Cross-aligned multimodal representation learning - modality specific representations and modality invariant representations learnt by separate networks and fused by a gating attention mechanism and concatenation. Survival function is cox partial likelihood loss.* | 629 | - | - | 240 | Hand crafted CellProfiler | - | Feature | Cross-validation | - | C-Statistic / AUC graph / Risk group curves | - | - | 0.841 | ±0.20 (unclear) | 0.889 | ±0.017 (unclear) |
| | Zhou, 2023 (66) | ***CMTA*** *Cross-modal translation alignment - two parallel encoder-decoder structures integrate intra-modal* | 569 | - | - | - | Learned ResNet-50 (ImageNet) | FC NN | Feature | Cross-validation | - | C-Statistic / Log-rank test / Risk group curves | - | - | 0.853 | ±0.0116 (unclear) | - | - |
| | | *representations and cross-modal representations generated by an attention mechanism.* | | | | | | | | | | | | | | | | |
| | *Zhou, 2023 (67)* | ***Integrative CNN** Tile level CNN image features concatenated with clinical variables and mutation status and/or signature. Final feature vector into MLP to predict risk.* | *-* | *-* | *-* | *-* | *Learned Inception V3 (ImageNet)* | *MSI status* | *Feature* | *Cross-validation* | *-* | *C-Statistic / AUC graph / Risk group curves* | *-* | *-* | *0.69* | *±0.19 (unclear)* | *0.8* | *x* |
| | *Zhu, 2023 (69)* | ***SAMMS** Pre-trained SAM for images + omic data into modality specific and cross-modality common subnetworks. Outputs concatenated with clinical data and fusion representation used for cox prediction layer.* | *500* | *-* | *-* | *102* | *Learned SAM (pretrained)* | *-* | *Feature* | *Cross-validation* | *-* | *C-Statistic / AUC graph / Log-rank test / Risk group curves* | *-* | *-* | *0.843* | *x* | *0.782* | *x* |
| | *Liu, 2024 (37)* | ***IntraSA-InterCA** Modality specific and intramodality modules for generation of feature embeddings integrated by an adaptive fusion block into a combined feature. Survival prediction via multilayer deep neural network.* | *345* | *-* | *22779* | *64* | *Hand crafted CellProfiler* | *-* | *Feature* | *Cross-validation* | *-* | *C-Statistic / AUC graph / Log-rank test / Risk group curves* | *Specificity, accuracy, sensitivity, precision, F1* | *-* | *0.742* | *±0.039 (unclear)* | *0.841* | *±0.032* |
| | *Perez-Herrera, 2024 (42)* | ***Model 3** Integration of image derived variables (UNet), omic features (multiple models) with clinical features via Cox Proportional Hazards model* | *3593* | *-* | *-* | *12* | *Learned -* | *-* | *Feature* | *Cross-validation* | *-* | *Hazard ratios* | *-* | *-* | *-* | *-* | *-* | *-* |
| | *Qiu, 2024 (43)* | ***DDM-Net** Dual-space disentangled-multimodal adversarial autoencoder comprising two visual autoencoders and a fusion module input to a cox proportional hazards model for survival prediction. Two imputation networks designed to handle missing data.* | *954* | *-* | *121415* | *2048* | *Learned ResNet50 (ImageNet)* | *SNN* | *Feature* | *Cross-validation* | *-* | *C-Statistic / Log-rank test / Risk group curves* | *-* | *-* | *0.824* | *±0.10 (unclear)* | *-* | *-* |
| | *Xie, 2024 (59)* | ***GaCaMML*** *Extracted features fed through cross-modal attention layer and MIL aggregation layer of model before fusion of gene and image features by concatenation. The log-likelihood loss function is used for survival prediction.* | *63* | *12 (19)* | *30000* | *256* | *Learned ResNet50 (ImageNet)* | *FC NN* | *Feature* | *Cross-validation* | *-* | *C-Statistic / Log-rank test / Risk group curves* | *-* | *-* | *0.613* | *x* | *-* | *-* |
| | *Zheng, 2024 (65)* | ***FSM*** *Undirected graph embeddings combined using an attention-based mechanism* | *444* | *156 (35)* | *-* | *-* | *Learned CNN (pretrained)* | *FC NN* | *Feature* | *Cross-validation* | *-* | *C-Statistic / AUC graph* | *-* | *-* | *0.703* | *0.017 (StD)* | *0.679* | *0.06 (StD)* |
| | *Zhou, 2024 (68)* | ***MSEN*** *Multimodal Survival Ensemble Network - unimodal bag representations of WSIs and omic data are passed through feature encoders and attention-based modules, then integrated via an ensemble strategy and input to Cox Proportional Hazard model* | *492* | *-* | *-* | *-* | *Learned HIPT* | *Ensemble based encoder* | *Feature* | *Cross-validation* | *-* | *C-Statistic / Risk group curves* | *-* | *-* | *0.827* | *±0.018 (StD)* | *-* | *-* |

**Abbreviations:** Calib. (Calibration), CI (Confidence Interval), Clin. (Clinical), CNN (convolutional neural network), DCA (Decision Curve Analysis), DCGMM (Deep conditional gaussian mixture model), FC (Fully Connected), GCN (Graph convolutional network), GSEA (Gene Set Enrichment Analysis), HIPT (Hierarchical Image Pyramid Transformer), HRD (Homologous Recombination Deficiency), KNN (K Nearest Neighbour), lmQCM (local maximal Quasi-Clique Merger),MLP (Multi-Layer Perceptron), PFTAS (Parameter Free Threshold Adjacency Statistics), SAM (Segment Anything Model), SGCN (Sparse Graph Convolutional Network), SNN (Self normalising network), StD (Standard Deviation), StErr (Standard Error), WGCNA (Weighted Gene Co-expression Network Analysis)

## Table 3: Risk of bias and applicability assessment

| Author, Year (ref) | Development | | | | | | | Evaluation | | | | | | | Overall | | | |
|---|---|---|---|---|---|---|---|---|---|---|---|---|---|---|---|---|---|---|
| | Quality | | | | Applicability | | | Risk of Bias | | | | Applicability | | | Dev | | Eval | |
| | 1. Participants | 2. Predictors | 3. Outcome | 4. Analysis | 1. Participants | 2. Prctrs | 3. Outcome | 1. Participants | 2. Predictors | 3. Outcome | 4. Analysis | 1. Participants | 2. Predictors | 3. Outcome | **Quality** | **Applicability** | **Risk of Bias** | **Applicability** |
| *Boehm, 2022 (22)* | ? | - | + | ? | + | ? | + | ? | - | + | - | + | ? | + | - | ? | - | ? |
| *Cheerla, 2019 (23)* | ? | ? | + | ? | + | + | + | ? | ? | + | - | + | + | + | ? | + | - | + |
| *Chen, 2021 (24)* | ? | ? | + | ? | + | + | + | ? | ? | + | ? | + | + | + | ? | + | ? | + |
| *Chen, 2021 (25)* | ? | ? | + | ? | + | + | + | ? | ? | + | - | + | + | + | ? | + | - | + |
| *Chen, 2022 (26)* | ? | ? | ? | - | + | + | ? | ? | ? | ? | - | + | + | ? | - | ? | - | ? |
| *Chen, 2022 (27)* | ? | ? | + | ? | + | + | ? | ? | ? | + | - | + | + | ? | ? | ? | - | ? |
| *Cheng, 2017 (28)* | ? | ? | + | - | + | + | + | ? | ? | + | - | + | + | + | - | + | - | + |
| *Hao, 2020 (29)* | ? | ? | ? | - | + | + | ? | ? | ? | ? | - | + | + | ? | - | ? | - | ? |
| *Hou, 2022 (30)* | ? | ? | + | - | + | + | + | ? | ? | + | - | + | + | + | - | + | - | + |
| *Hou, 2023 (31)* | ? | ? | + | ? | + | + | ? | ? | ? | + | - | + | + | ? | ? | ? | - | ? |
| *Ji, 2024 (32)* | ? | ? | + | - | + | + | + | ? | ? | + | ? | + | + | + | - | + | ? | + |
| *Li, 2020 (33)* | ? | ? | ? | ? | + | + | ? | ? | ? | ? | - | + | + | ? | ? | ? | - | ? |
| *Li, 2021 (34)* | ? | ? | + | ? | + | + | + | ? | ? | + | ? | + | + | + | ? | + | ? | + |
| *Li, 2022 (35)* | ? | ? | + | - | + | + | ? | ? | ? | + | - | + | + | ? | - | ? | - | ? |
| *Liu, 2023 (36)* | ? | ? | + | ? | + | + | + | ? | ? | + | - | + | + | + | ? | + | - | + |
| *Liu, 2024 (37)* | ? | ? | + | - | + | + | ? | ? | ? | + | - | + | + | ? | - | ? | - | ? |
| *Lv, 2021 (38)* | ? | ? | + | ? | + | + | ? | ? | ? | + | - | + | + | ? | ? | ? | - | ? |
| *Lv, 2023 (39)* | ? | ? | + | ? | + | + | ? | ? | ? | + | - | + | + | ? | ? | ? | - | ? |
| *Ning, 2020 (40)* | ? | ? | + | - | + | + | ? | ? | ? | + | - | + | + | ? | - | ? | - | ? |
| *Ning, 2023 (41)* | ? | ? | + | - | + | + | ? | ? | ? | + | - | + | + | ? | - | ? | - | ? |
| *Perez-Herrera, 2024 (42)* | ? | ? | + | ? | + | + | ? | ? | ? | + | - | + | + | ? | ? | ? | - | ? |
| *Qiu, 2024 (43)* | ? | ? | + | ? | + | + | ? | ? | ? | + | - | + | + | ? | ? | ? | - | ? |
| *Shao, 2020 (44)* | ? | ? | + | - | + | + | ? | ? | ? | + | - | + | + | ? | - | ? | - | ? |
| *Shao, 2023 (45)* | ? | ? | + | - | + | + | ? | ? | ? | + | - | + | + | ? | - | ? | - | ? |
| *Shao, 2020 (46)* | ? | ? | + | - | + | + | ? | ? | ? | + | - | + | + | ? | - | ? | - | ? |
| *Shao, 2023 (47)* | ? | ? | + | - | + | + | ? | ? | ? | + | - | + | + | ? | - | ? | - | ? |
| *Steyaert, 2023 (48)* | ? | + | + | - | + | + | + | ? | + | + | - | + | + | + | - | + | - | + |
| *Subramanian, 2021 (49)* | ? | ? | + | ? | + | + | ? | ? | ? | + | - | + | + | ? | ? | ? | - | ? |
| *Subramanian, 2024 (50)* | ? | ? | + | - | + | + | + | ? | ? | + | - | + | + | + | - | + | - | + |
| *Sun, 2018 (51)* | ? | ? | + | ? | + | + | ? | ? | ? | + | - | + | + | ? | ? | ? | - | ? |
| *Tan, 2022 (52)* | ? | ? | + | - | + | + | ? | ? | ? | + | - | + | + | ? | - | ? | - | ? |
| *Vale-Silva, 2021 (53)* | ? | ? | + | ? | + | + | ? | ? | ? | + | - | + | + | ? | ? | ? | - | ? |
| *Vollmer, 2024 (54)* | - | ? | + | ? | + | + | + | - | ? | + | - | + | + | + | - | + | - | + |
| *Wang, 2021 (55)* | ? | ? | + | - | + | + | ? | ? | ? | + | - | + | + | ? | - | ? | - | ? |
| *Wang, 2023 (56)* | ? | ? | + | - | + | + | ? | ? | ? | + | - | + | + | ? | - | ? | - | ? |
| *Wei, 2023 (57)* | ? | ? | + | ? | + | + | + | ? | ? | + | - | + | + | + | ? | + | - | + |
| *Wu, 2023 (58)* | ? | ? | + | - | + | + | ? | ? | ? | + | - | + | + | ? | - | ? | - | ? |
| *Xie, 2024 (59)* | ? | ? | + | - | + | + | ? | ? | ? | + | - | + | + | ? | - | ? | - | ? |
| *Zeng, 2020 (60)* | ? | ? | + | ? | + | + | + | ? | ? | + | - | + | + | + | ? | + | - | + |
| *Zeng, 2021 (61)* | ? | ? | + | - | + | + | + | ? | ? | + | - | + | + | + | - | + | - | + |
| *Zhan, 2021 (62)* | ? | ? | + | - | + | + | ? | ? | ? | + | - | + | + | ? | - | ? | - | ? |
| *Zhang, 2020 (63)* | ? | ? | + | - | + | + | ? | ? | ? | + | - | + | + | ? | - | ? | - | ? |
| *Zhao, 2023 (64)* | ? | ? | + | - | + | + | ? | ? | ? | + | - | + | + | ? | - | ? | - | ? |
| *Zheng, 2024 (65)* | ? | ? | + | ? | + | + | ? | ? | ? | + | - | + | + | ? | ? | ? | - | ? |
| *Zhou, 2023 (66)* | ? | ? | + | ? | + | + | + | ? | ? | + | - | + | + | + | ? | + | - | + |
| *Zhou, 2023 (67)* | ? | - | + | ? | + | + | + | ? | - | + | - | + | + | + | - | + | - | + |
| *Zhou, 2024 (68)* | ? | ? | + | ? | + | + | ? | ? | ? | + | - | + | + | ? | ? | ? | - | ? |
| *Zhu, 2023 (69)* | ? | ? | + | - | + | + | ? | ? | ? | + | - | + | + | ? | - | ? | - | ? |

## Supplementary materials

## Literature search strategy

Search strategies for the three databases shown below with numbers of results for each indicated.

**1. PUBMED (12/08/2024)**

| **Search** | **Term [title/abstract]** | **Number** |
|---|---|---|
| *Pathology whole slide images* | | |
| #1 | Pathology | 427135 |
| #2 | Histo* | 2238976 |
| #3 | “H&e” | 19770 |
| #4 | “Haematoxylin and eosin” | 3688 |
| #5 | Imag* | 751150 |
| #6 | “digital pathology” | 2714 |
| #7 | “whole slide image*” | 2192 |
| #8 | “WSIs” | 863 |
| #9 | ((#1 OR #2 OR #3 OR #4) AND #5) OR # 6 OR #7 OR #8 | 89536 |
| *Omic data* | | |
| #10 | Genom* | 874583 |
| #11 | Omic* | 44061 |
| #12 | Proteom* | 146305 |
| #13 | Epigen* | 135526 |
| #14 | Transcriptom* | 166893 |
| #15 | Molecular | 1828450 |
| #16 | #10 OR #11 OR #12 OR #13 OR #14 OR #15 | 27992170 |
| *Pathogenomic* | | |
| #17 | Pathogenom* | 268 |
| #18 | (#9 AND #16) OR #17 | 4932 |
| *Machine learning method* | | |
| #19 | “deep learning” | 68880 |
| #20 | “machine learning” | 121049 |
| #21 | “neural network” | 80989 |
| #22 | “artificial intelligence” | 55905 |
| #23 | Fusion | 255302 |
| #24 | Integrat* | 792949 |
| #25 | Multimodal | 67428 |
| #26 | multi-modal | 7934 |
| #27 | (#19 OR #20 OR #21 OR #22) AND (#23 OR #24 OR #25 OR #26) | 39060 |
| *Cancer* | | |
| #28 | Cancer* | 245687 |
| #29 | Tumo?r* | 2157883 |
| #30 | Carcinoma* | 815576 |
| #31 | Sarcoma* | 118675 |
| #32 | Onco* | 449173 |
| #33 | #28 Or #29 OR #30 OR #31 OR #32 | 4062912 |
| ***Final search*** | | |
| #33 | #18 AND #27 AND #33 | 215 |
| #33 | English, human | 128 |

**2. OVID: Embase classic+ Embase, MEDLINE (R) ALL (12/08/2024)**

| Search | Term [title/abstract] | Number |
|---|---|---|
| *Pathology whole slide images* | | |
| #1 | Pathology | 964013 |
| #2 | Histo* | 5508575 |
| #3 | "H&e" | 718077 |
| #4 | "Haematoxylin and eosin" | 13620 |
| #5 | Imag* | 1753707 |
| #6 | "digital pathology" | 5003 |
| #7 | "whole slide image*" | 5487 |
| #8 | "WSIs" | 2156 |
| #9 | ((#1 OR #2 OR #3 OR #4) AND #5) OR # 6 OR #7 OR #8 | 247780 |
| *Omic data* | | |
| #10 | Genom* | 1902327 |
| #11 | Omic* | 92389 |
| #12 | Proteom* | 322280 |
| #13 | Epigen* | 300859 |
| #14 | Transcriptom* | 365122 |
| #15 | Molecular | 3944775 |
| #16 | #10 OR #11 OR #12 OR #13 OR #14 OR #15 | 6054934 |
| *Pathogenomic* | | |
| #17 | Pathogenom* | 501 |
| #18 | (#9 AND #16) OR #17 | 14278 |
| *Machine learning method* | | |
| #19 | "deep learning" | 128669 |
| #20 | "machine learning" | 231297 |
| #21 | "neural network" | 163613 |
| #22 | "artificial intelligence" | 102044 |
| #23 | Fusion | 567992 |
| #24 | Integrat* | 17333754 |
| #25 | Multimodal | 138719 |
| #26 | multi-modal | 19255 |
| #27 | (#19 OR #20 OR #21 OR #22) AND (#23 OR #24 OR #25 OR #26) | 76067 |
| *Cancer* | | |
| #28 | Cancer* | 5827011 |
| #29 | Tumo?r* | 5190000 |
| #30 | Carcinoma* | 1942422 |
| #31 | Sarcoma* | 274519 |
| #32 | Onco* | 1121283 |
| #33 | #28 Or #29 OR #30 OR #31 OR #32 | 9712811 |
| ***Final search*** | | |
| #33 | #18 AND #27 AND #33 | 528 |
| #33 | Limit [English language] | 522 |
| #34 | Limit [humans] | 433 |
| #35 | Remove duplicates from #34 | 320 |

**3. CENTRAL**

| **Search** | **Term [title/abstract/keyword]** | **Number** |
|---|---|---|

*Pathology whole slide images*

| | | |
|---|---|---|
| #1 | Pathology | 83423 |
| #2 | Histo* | 150560 |
| #3 | “H&E” | 701 |
| #4 | “Haematoxylin and eosin” | 386 |
| #5 | Imag* | 40806 |
| #6 | “digital pathology” | 82 |
| #7 | Whole NEXT slide NEXT image* | 73 |
| #8 | “WSIs” | 16 |
| #9 | ((#1 OR #2 OR #3 OR #4) AND #5) OR # 6 OR #7 OR #8 | 9659 |

*Omic Data*

| | | |
|---|---|---|
| #10 | Genom* | 9172 |
| #11 | Omic* | 927 |
| #12 | Proteom* | 2021 |
| #13 | Epigen* | 1846 |
| #14 | Transcriptom* | 2042 |
| #15 | Molecular | 30428 |
| #16 | #10 OR #11 OR #12 OR #13 OR #14 OR #15 | 42243 |

*Pathogenomic*

| | | |
|---|---|---|
| #17 | Pathogenom* | 1 |
| #18 | (#9 AND #16) OR #17 | 515 |

*Machine learning method*

| | | |
|---|---|---|
| #19 | “deep learning” | 1207 |
| #20 | “machine learning” | 3108 |
| #21 | “neural network” | 1525 |
| #22 | “artificial intelligence” | 2388 |
| #23 | Fusion | 10218 |
| #24 | Integrat* | 48667 |
| #25 | Multimodal | 9740 |
| #26 | multi-modal | 1189 |
| #27 | (#19 OR #20 OR #21 OR #22) AND (#23 OR #24 OR #25 OR #26) | 784 |
| *Cancer* | | |
| #28 | Cancer* | 238625 |
| #29 | Tumo?r* | 100558 |
| #30 | Carcinoma* | 54398 |
| #31 | Sarcoma* | 3502 |
| #32 | Onco* | 104041 |
| #33 | #28 Or #29 OR #30 OR #31 OR #32 | 291641 |
| ***Final search*** | | |
| #33 | #18 AND #27 AND #33 | 15 |
| #33 | In trials | 9 |

## 1.7 Screening algorithms

### 1.7.1 Abstract screening

| Signalling questions for inclusion and exclusion of studies in abstract screening | | |
|---|---|---|
| **#** | **Questions** | **Response & Action** |
| 1 | Is this study written in English? | No = Reject "not English"<br>Yes = Next question |
| 2 | Is this article a primary research paper?<br>(ie. not a meta-analysis, review, comment, conference abstract, book chapter etc) | No = Reject "wrong publication type"<br>Yes = Next question |
| 3 | Is this article examining cancer?<br>(ie not benign tumor, dysplasia or other non-cancerous medical conditions) | No = Reject "not cancer"<br>Yes = Next question |
| 4 | Are the participants human?<br>(ie. Not cell culture or animal study) | No = Reject "not human"<br>Yes = Next question |
| 5 | Is the study concerned with a novel prognostic/prediction model?<br>(ie.not a model developed to predict an existing prognostic factor eg. BRAF status from the images, or only identifying new prognostic factors) | No = Reject "not prognostic"<br>Yes = Next question |
| 6 | Is the prognostic outcome overall survival or cancer-specific survival?<br>(exclude disease free progression, risk of recurrence, treatment response predictions) | No = reject "wrong outcome"<br>Yes = next question |
| 7 | Was the model developed through machine learning methods?<br>(ie. Not other standard statistical methods)<br>Note: logistic regression accepted as machine learning IF described as such by authors. | No = Reject "not ML"<br>Yes = Next question |
| 8 | Are whole slide images of surgical pathology part of the model input?<br>IHC and special stains included, TMAs<br>(Not other imaging modality eg. other pathology imaging technologies such as multiplex IHC or immunofluorescence images, radiological imaging, endoscopy etc AND Not cytology, autopsy, toxicology, forensics) | No = Reject "no WSIs"<br>Yes = Next question |
| 9 | Is high throughput molecular/omic data part of the model input?<br>(ie.genomes, transcriptomes, proteomes, epigenetic data but NOT simple point mutation testing outcomes eg. BRAFV600E wt/mutant).<br>Note: Spatial transcriptomic data using multiplex imaging accepted where light microscope WSI's also included but not as the only image modality. | No = Reject "no omics"<br>Yes = Next question |

## 1.7.2 Full text screening

| Signalling questions for inclusion and exclusion of studies in full-text screening | | |
|---|---|---|
| **#** | **Question** | **Response & Action** |
| 1 | Is this study written in English? | No = Reject "not English"<br>Yes = Next question |
| 2 | Is this article a primary research paper?<br>(ie. not a meta-analysis, review, comment, conference abstract, book chapter etc) | No = Reject "wrong publication type"<br>Yes = Next question |
| 3 | Is this article peer reviewed?<br>Ie. Not pre-print or other type of non-peer reviewed article | No = reject "wrong publication type"<br>Yes = Next question |
| 4 | Is this article examining cancer?<br>(ie not benign tumor, dysplasia or other non-cancerous medical conditions) | No = Reject "not cancer"<br>Yes = Next question |
| 5 | Are the participants human?<br>(ie. Not cell culture or animal study) | No = Reject "not human"<br>Yes = Next question |
| 6 | Is the study concerned with a novel prognostic/prediction model?<br>(ie.not a model developed to predict an existing prognostic factor eg. BRAF status from the images, or only identifying new prognostic factors) | No = Reject "not prognostic"<br>Yes = Next question |
| 7 | Is the prognostic outcome overall survival or cancer-specific survival?<br>(exclude disease free progression, risk of recurrence, treatment response predictions) | No = reject "wrong outcome"<br>Yes = next question |
| 8 | Was the model developed through machine learning methods?<br>(ie. Not other standard statistical methods)<br>Note: logistic regression accepted as machine learning IF described as such by authors. | No = Reject "not ML"<br>Yes = Next question |
| 9 | Are whole slide images of surgical pathology part of the model input?<br>IHC and special stains included, TMAs<br>(Not other imaging modality eg. other pathology imaging technologies such as multiplex IHC or immunofluorescence images, radiological imaging, endoscopy etc AND Not cytology, autopsy, toxicology, forensics) | No = Reject "no WSIs"<br>Yes = Next question |
| 10 | Is high throughput molecular/omic data part of the model input?<br>(ie.genomes, transcriptomes, proteomes, epigenetic data but NOT simple point mutation testing outcomes eg. BRAFV600E wt/mutant).<br>Note: Spatial transcriptomic data using multiplex imaging accepted where light microscope WSI's also included but not as the only image modality. | No = Reject "no omics"<br>Yes = Next question |
| 11 | Are these data modalities integrated into a single prognostic model? | No = Reject "not integrated"<br>Yes = include |

## 1.8 Data extraction template summary

The table below summarises the data extraction fields for the systematic review.

The template used was adapted from Fernandex-Felix et al, 2023.

Fernandez-Felix, B.M., López-Alcalde, J., Roqué, M. *et al.* CHARMS and PROBAST at your fingertips: a template for data extraction and risk of bias assessment in systematic reviews of predictive models. *BMC Med Res Methodol* **23**, 44 (2023). https://doi.org/10.1186/s12874-023-01849-0

Adaptations included updating for PROBAST+AI and to enable additional domain specific fields, fields relating specifically to multimodal fusion methods and open science reporting fields for data and code (highlighted in bold). Please see the original publication to access the template.

| **Summary info** | |
|---|---|
| Author; Lead author country; publication year; study title; journal; model name; cancer domain (organ); cancer subtype | |
| **Study info** | |
| Source of data | Source of data (retrospective, prospective etc); name of dataset; country of origin; **WSI details (fixation, stain, resolution, scanning platform); omic data types (RNA, miRNA etc)** |
| Participants | Recruitment method; recruitment dates; study setting; study sites (regions); study sites (no. centres); inclusion criteria; exclusion criteria; participant description; Numbers (participants, **WSI, genes**); participant characteristics (**demographic information available (list); disease information available (list); treatment info provided (list)**) |
| Outcome | Outcome; outcome definition; type of outcome (single, combined); assessed without knowledge predictors (y/n); |
| | predictors part of outcome (y/n); time of outcome occurrence |
| Candidate predictors | Number assessed; type; timing of measurement; predictor measurement similar for all (y/n); predictors blinded for outcome (y/n); handling continuous predictors. |
| Sample size | Number participants; number events; epv/epp |
| Missing data | Number participants with missing data; handling of missing data |
| Model development | Modelling method; **method for generating WSI predictors (pre-processing, sampling resolution, patch size, patch number, patch selection, image feature generation, no. features);** selection WSI predictors; **omic predictor generation (pre-processing, handling missing values, gene feature generation, no. gene features);** selection of omic predictors. **Method multimodal fusion (feature, decision, combination**); method selection during modelling' methods to prevent overfitting |
| Model performance | Calibration; discrimination; overall; clinical utility |
| Model evaluation | Internal validation; external validation; model comparison (unimodal, alt multimodal) |
| Results | Final no predictors in model; alternative presentation of model |
| Interpretation | Explainability/feature importance methods |
| Observations | **Data availability; code availability; funding** |

## 1.9 Author definitions

### 1.9.1 Model Categories used in the review

1. **Regularised Regression Models (e.g., LASSO-Cox, Elastic Net)**

Definition: Models using Cox or logistic regression with machine learning-based regularisation to improve generalisation and automate feature selection.

2. **Classical Machine Learning Models (non-DL)**

Definition: Random forests, support vector machines, gradient boosting, k-nearest neighbors, etc., using structured/hand-crafted features.

3. **Deep Learning Models**

Definition: CNNs, autoencoders, graph networks, transformer and foundation model based approaches

### 1.9.2 Fusion approach definitions

**Feature fusion: Combining features from separate data modalities into a shared representation to generate model predictions.**

**Decision Fusion:** Predictions or decisions from separate unimodal models are combined at a late stage, typically after each modality has been fully processed and independently classified or regressed, to produce a final output.

**Hybrid:** Combines elements of feature and decision fusion, such as integrating the decision-level output (e.g., risk score) from one modality with the feature-level representation of another.

## 1.10 PRISMA 2020 checklists

Checklists completed and generated using https://prisma.shinyapps.io/checklist/

### 1.10.1 PRISMA 2020 Main Checklist

| Topic | No. | Item | Location where item is reported |
|---|---|---|---|
| **TITLE** | | | |
| **Title** | 1 | Identify the report as a systematic review. | Page 1, title |
| **ABSTRACT** | | | |
| **Abstract** | 2 | See the PRISMA 2020 for Abstracts checklist | |
| **INTRODUCTION** | | | |

| Topic | No. | Item | Location where item is reported |
|---|---|---|---|
| **Rationale** | 3 | Describe the rationale for the review in the context of existing knowledge. | Introduction, page 3-5 |
| **Objectives** | 4 | Provide an explicit statement of the objective(s) or question(s) the review addresses. | Introduction, page 5 |
| **METHODS** | | | |
| **Eligibility criteria** | 5 | Specify the inclusion and exclusion criteria for the review and how studies were grouped for the syntheses. | Methods; eligibility criteria, page 6 |
| **Information sources** | 6 | Specify all databases, registers, websites, organisations, reference lists and other sources searched or consulted to identify studies. Specify the date when each source was last searched or consulted. | Methods; data sources and search strategy, page 6 |
| **Search strategy** | 7 | Present the full search strategies for all databases, registers and websites, including any filters and limits used. | Supplementary file; Literature search strategy, page 1-6 |
| **Selection process** | 8 | Specify the methods used to decide whether a study met the inclusion criteria of the review, including how many reviewers screened each record and each report retrieved, whether they worked independently, and if applicable, details of automation tools used in the process. | Methods; study selection, page 6-7 & Supplementary file; screening algorithms, page 6-9 |

| Topic | No. | Item | Location where item is reported |
|---|---|---|---|
| **Data collection process** | 9 | Specify the methods used to collect data from reports, including how many reviewers collected data from each report, whether they worked independently, any processes for obtaining or confirming data from study investigators, and if applicable, details of automation tools used in the process. | Methods; data extraction, page 7-8 |
| **Data items** | 10a | List and define all outcomes for which data were sought. Specify whether all results that were compatible with each outcome domain in each study were sought (e.g. for all measures, time points, analyses), and if not, the methods used to decide which results to collect. | Methods; data extraction, page 7 & Supplementary file; data extraction template summary, page 8-9 |
| | 10b | List and define all other variables for which data were sought (e.g. participant and intervention characteristics, funding sources). Describe any assumptions made about any missing or unclear information. | Supplementary file; data extraction summary, page 8-9 |
| **Study risk of bias assessment** | 11 | Specify the methods used to assess risk of bias in the included studies, including details of the tool(s) used, how many reviewers assessed each study and whether they worked independently, and if applicable, details of automation tools used in the process. | Methods; risk of bias assessment, page 8-9 |

| Topic | No. | Item | Location where item is reported |
|---|---|---|---|
| **Effect measures** | 12 | Specify for each outcome the effect measure(s) (e.g. risk ratio, mean difference) used in the synthesis or presentation of results. | Methods; data synthesis, page 9 |
| **Synthesis methods** | 13a | Describe the processes used to decide which studies were eligible for each synthesis (e.g. tabulating the study intervention characteristics and comparing against the planned groups for each synthesis (item 5)). | Methods; data synthesis, page 9 |
| | 13b | Describe any methods required to prepare the data for presentation or synthesis, such as handling of missing summary statistics, or data conversions. | N/A |
| | 13c | Describe any methods used to tabulate or visually display results of individual studies and syntheses. | Methods; data synthesis, page 9 |
| | 13d | Describe any methods used to synthesize results and provide a rationale for the choice(s). If meta-analysis was performed, describe the model(s), method(s) to identify the presence and extent of statistical heterogeneity, and software package(s) used. | Methods; data synthesis, page 9 |
| | 13e | Describe any methods used to explore possible causes of heterogeneity among study results (e.g. subgroup analysis, meta-regression). | N/A |

| Topic | No. | Item | Location where item is reported |
|---|---|---|---|
| | 13f | Describe any sensitivity analyses conducted to assess robustness of the synthesized results. | N/A |
| **Reporting bias assessment** | 14 | Describe any methods used to assess risk of bias due to missing results in a synthesis (arising from reporting biases). | N/A |
| **Certainty assessment** | 15 | Describe any methods used to assess certainty (or confidence) in the body of evidence for an outcome. | N/A |
| **RESULTS** | | | |
| **Study selection** | 16a | Describe the results of the search and selection process, from the number of records identified in the search to the number of studies included in the review, ideally using a flow diagram. | Results, page 9-10 & Figure 1, prism flow diagram |
| | 16b | Cite studies that might appear to meet the inclusion criteria, but which were excluded, and explain why they were excluded. | Results, page 10 |
| **Study characteristics** | 17 | Cite each included study and present its characteristics. | Table 1 |
| **Risk of bias in studies** | 18 | Present assessments of risk of bias for each included study. | Table 3 |

| Topic | No. | Item | Location where item is reported |
|---|---|---|---|
| **Results of individual studies** | 19 | For all outcomes, present, for each study: (a) summary statistics for each group (where appropriate) and (b) an effect estimate and its precision (e.g. confidence/credible interval), ideally using structured tables or plots. | Table 2 |
| **Results of syntheses** | 20a | For each synthesis, briefly summarise the characteristics and risk of bias among contributing studies. | N/A |
| | 20b | Present results of all statistical syntheses conducted. If meta-analysis was done, present for each the summary estimate and its precision (e.g. confidence/credible interval) and measures of statistical heterogeneity. If comparing groups, describe the direction of the effect. | N/A |
| | 20c | Present results of all investigations of possible causes of heterogeneity among study results. | N/A |
| | 20d | Present results of all sensitivity analyses conducted to assess the robustness of the synthesized results. | N/A |
| **Reporting biases** | 21 | Present assessments of risk of bias due to missing results (arising from reporting biases) for each synthesis assessed. | N/A |
| **Certainty of evidence** | 22 | Present assessments of certainty (or confidence) in the body of evidence for each outcome assessed. | N/A |

| Topic | No. | Item | Location where item is reported |
|---|---|---|---|
| **DISCUSSION** | | | |
| **Discussion** | 23a | Provide a general interpretation of the results in the context of other evidence. | Discussion, page 17-19 |
| | 23b | Discuss any limitations of the evidence included in the review. | Discussion, page 17-19 |
| | 23c | Discuss any limitations of the review processes used. | Discussion; limitations of the review, page 19-20 |
| | 23d | Discuss implications of the results for practice, policy, and future research. | Discussion, current limitations and future recommendations, page 20-21 |
| **OTHER INFORMATION** | | | |
| **Registration and protocol** | 24a | Provide registration information for the review, including register name and registration number, or state that the review was not registered. | Methods, page 5 |
| | 24b | Indicate where the review protocol can be accessed, or state that a protocol was not prepared. | Methods, page 5 |
| | 24c | Describe and explain any amendments to information provided at registration or in the protocol. | N/A |
| **Support** | 25 | Describe sources of financial or non-financial support for the review, and the role of the funders or sponsors in the review. | Funding, page 21 |

| Topic | No. | Item | Location where item is reported |
|---|---|---|---|
| **Competing interests** | 26 | Declare any competing interests of review authors. | Competing interests statement, page 22 |
| **Availability of data, code and other materials** | 27 | Report which of the following are publicly available and where they can be found: template data collection forms; data extracted from included studies; data used for all analyses; analytic code; any other materials used in the review. | Data availability statement, page 22 |

## 1.10.2 PRISMA Abstract Checklist

| Topic | No. | Item | Reported? |
|---|---|---|---|
| **TITLE** | | | |
| **Title** | 1 | Identify the report as a systematic review. | Yes |
| **BACKGROUND** | | | |
| **Objectives** | 2 | Provide an explicit statement of the main objective(s) or question(s) the review addresses. | Yes |
| **METHODS** | | | |
| **Eligibility criteria** | 3 | Specify the inclusion and exclusion criteria for the review. | Yes |
| **Information sources** | 4 | Specify the information sources (e.g. databases, registers) used to identify studies and the date when each was last searched. | Yes |
| **Risk of bias** | 5 | Specify the methods used to assess risk of bias in the included studies. | Yes |
| **Synthesis of results** | 6 | Specify the methods used to present and synthesize results. | Yes |
| **RESULTS** | | | |

| Topic | No. | Item | Reported? |
|---|---|---|---|
| **Included studies** | 7 | Give the total number of included studies and participants and summarise relevant characteristics of studies. | Yes |
| **Synthesis of results** | 8 | Present results for main outcomes, preferably indicating the number of included studies and participants for each. If meta-analysis was done, report the summary estimate and confidence/credible interval. If comparing groups, indicate the direction of the effect (i.e. which group is favoured). | Yes |
| **DISCUSSION** | | | |
| **Limitations of evidence** | 9 | Provide a brief summary of the limitations of the evidence included in the review (e.g. study risk of bias, inconsistency and imprecision). | Yes |
| **Interpretation** | 10 | Provide a general interpretation of the results and important implications. | Yes |
| **OTHER** | | | |
| **Funding** | 11 | Specify the primary source of funding for the review. | Yes |
| **Registration** | 12 | Provide the register name and registration number. | Yes |

*From:* Page MJ, McKenzie JE, Bossuyt PM, Boutron I, Hoffmann TC, Mulrow CD, et al. The PRISMA 2020 statement: an updated guideline for reporting systematic reviews. MetaArXiv. 2020, September 14. DOI: 10.31222/osf.io/v7gm2. For more information, visit: www.prisma-statement.org

### 1.10.3 SWiM 2021 checklist

Meta-analysis explanation and elaboration article is: Campbell M, McKenzie JE, Sowden A, Katikireddi SV, Brennan SE, Ellis S, Hartmann-Boyce J, Ryan R, Shepperd S, Thomas J, Welch V, Thomson H. Synthesis without meta-analysis (SWiM) in systematic reviews: reporting guideline BMJ 2020;368:l6890 http://dx.doi.org/10.1136/bmj.l6890

| **SWiM is intended to complement and be used as an extension to PRISMA** | | | |
|---|---|---|---|
| **SWiM reporting item** | **Item description** | **Page in manuscript where item is reported** | **Other*** |
| *Methods* | | | |
| **1** Grouping studies for synthesis | 1a) Provide a description of, and rationale for, the groups used in the synthesis (e.g., groupings of populations, interventions, outcomes, study design) | Page 9, "Data synthesis" | |
| | 1b) Detail and provide rationale for any changes made subsequent to the protocol in the groups used in the synthesis | N/A | |
| **2** Describe the standardised metric and transformation methods used | Describe the standardised metric for each outcome. Explain why the metric(s) was chosen, and describe any methods used to transform the intervention effects, as reported in the study, to the standardised metric, citing any methodological guidance consulted | Page 9, "Data synthesis: | |
| **3** Describe the synthesis methods | Describe and justify the methods used to synthesise the effects for each outcome when it was not possible to undertake a meta-analysis of effect estimates | Page 9, "Data synthesis" | |
| **4** Criteria used to prioritise results for summary and synthesis | Where applicable, provide the criteria used, with supporting justification, to select the particular studies, or a particular study, for the main synthesis or to draw conclusions from the synthesis (e.g., based on study design, risk of bias assessments, directness in relation to the review question) | Page 9, "Data synthesis" | |
| **SWiM reporting item** | **Item description** | **Page in manuscript where item is reported** | **Other*** |
| **5** Investigation of heterogeneity in reported effects | State the method(s) used to examine heterogeneity in reported effects when it was not possible to undertake a meta-analysis of effect estimates and its extensions to investigate heterogeneity | N/A | |
| **6** Certainty of evidence | Describe the methods used to assess certainty of the synthesis findings | N/A | |
| **7** Data presentation methods | Describe the graphical and tabular methods used to present the effects (e.g., tables, forest plots, harvest plots).<br>Specify key study characteristics (e.g., study design, risk of bias) used to order the studies, in the text and any tables or graphs, clearly referencing the studies included | Page 9, "Data synthesis" | |
| *Results* | | | |
| **8** Reporting results | For each comparison and outcome, provide a description of the synthesised findings, and the certainty of the findings. Describe the result in language that is consistent with the question the synthesis addresses, and indicate which studies contribute to the synthesis | Page 9-17, "Results" | |
| *Discussion* | | | |
| **9** Limitations of the synthesis | Report the limitations of the synthesis methods used and/or the groupings used in the synthesis, and how these affect the conclusions that can be drawn in relation to the original review question | Page 17, performance comparisons. Supplementary Figures 1-4. Page 18, "Limitations of the review" | |

PRISMA=Preferred Reporting Items for Systematic Reviews and Meta-Analyses.
*If the information is not provided in the systematic review, give details of where this information is available (e.g., protocol, other published papers (provide citation details), or website (provide the URL)).

## 1.11 Additional results

### 1.11.1 Table S1. Data processing

Table S1. Supplementary study and data processing details of the models of interest

| Author, Year (ref) | Model name (Key dataset) | WSI Pre-processing | Sampling resolution | Patch size | Patch selection | Patches per patient | WSI feature extraction | WSI feature extraction | Selection WSI fetaures | Omic Pre-processing | Omic features | Selection omic features | Intermodality interactions captured by method | Internal validation | | | External results | | Alternative modelling in study: | | | Comparisons: | |
|---|---|---|---|---|---|---|---|---|---|---|---|---|---|---|---|---|---|---|---|---|---|---|---|
| | | | | | | | Approach | Method | | | Named Method | | | Approach | Data split | No. folds | Result (metric) | Variability (metric) | Model approach | Data combo | Cancer dataset | Compare unimodal | Alt datset results |
| Boehm, 2022 (22) | **GHC***, (Ovary) HGSOC | Macenko Stain normalisation | 20x* | 128x128 | - | - | Combination | Tissue classification (ResNet18) and nuclear detection maps | Univariate CPH | Removal of cases with ambiguous HRD status | HRD status inferred from key genes associated | Univariate CPH | No | Random split data | 85:15 | N/A | - | - | - | Yes | - | Yes | - |
| Cheerla, 2019 (23) | **Multimodal (Clin+miRNA+mRNA+WSI)**, (Multi - 20) Pancancer | Pre-processed, batch corrected | - | 224x224 | Select top 20% closest to RGB mean | 40 | Learned | SqueezeNet CNN | - | - | Deep Highway Networks | All | Via unsupervised similarity loss, and multimodal dropout during training | Random split data | 85:15 | N/A | - | - | - | Yes | Yes | - | Yes |
| Chen, 2021 (24) | **HTRS**, (Head & Neck) SCC | Correct illumination | 40x | 1000x1000 | - | 20 | Hand crafted | CellProfiler | Overlap of SVM-RFE and LASSO-COX selection | - | WGCNA | Genes associated with image features | Partial ly -WGCNA identifies gene modules correlated with image features but no joint learned latent space | Cross-validation | 70:30 | 5 | - | - | - | - | - | Yes | - |
| Chen, 2021 (25) | **Multiomics model** , (Lung) Adenocarcinoma | Correct illumination | 40x* | 1000x1000 | Random | 60 | Hand crafted | CellProfiler | All | - | - | All proteomic; top 100 somatic mutations, top 100 DEG | No | Cross-validation | 50:50 | 5 | - | - | - | Yes | - | Yes | - |
| Chen, 2022 (26) | **MMF**, (Multi - 14) Pancancer | CLAM tissue segmentation | 20x | 256x256 | From identified tissue regions (unknown number) | - | Learned | ResNet50 (ImageNet) | All | Filter for Genes with >10% CNV or 5% mutation frequency, RNASeq-gene sets | - | - | Shared embedding with attention | Cross-validation | 80:20 | 5 | - | - | - | - | Yes | Yes | Yes |
| Chen, 2022 (27) | **Pathomic Fusion**, (Multi - 2) Glioma | Sparse stain normalisation | 20x | 512x512, 1024x1024 | - | - | Combination | CNN VGG19 (ImageNet), KNN-Cell graph (concatenation of manual, contrastive predictive coding features) | - | - | SNN | - | Gating-based attention + Kronecker product | Cross-validation | 80:20 | 15 (Monte carlo) | - | - | - | Yes | Yes | Yes | Yes |
| Cheng, 2017 (28) | **Lasso-Cox**, (Multi - 2) CCRCC, Glioma | Unsupervised hierarchical multilevel thresholding | - | - | - | - | Hand crafted | Unclear | - | Normalised to reads per kilobase per million | lmQCM | - | No | Cross-validation | - | 10 | - | - | - | - | Yes | - | Yes |
| Hao, 2020 (29) | **PAGE-NET**, (Brain) Glioblastoma | - | - | 256x256 | Random | 1000 | Learned | Pre-trained CNN | All | - | CoxPASNET | KEGG and Reactome datases to filter with biological knowledge | Attention-based fusion | Cross-validation | 90:10 | - | - | - | - | Yes | - | Yes | - |
| Hou, 2022 (30) | **Multi-modality**, (Liver) HCC | Orientation augmentation | 20x | 1024x1024 | - | 100-4000 | Learned | VGG19 | K means clustering | Missing values removed | genes greatest variance selected then WGCNA | Key modules selected by LASSO, top hub gene of 5 modules | No | Cross-validation | 67:33 | 5 | - | - | - | - | - | Yes | - |
| Hou, 2023 (31) | **HGCN**, (Multi - 6) KIRC | - | 10x | 512x512 | - | - | Learned | KimiaNet (CNN) | All | - | Gene set enrichment analysis | Genomic embeddings in functional groups | Hybrid graph convolutional networks (GCNs) and hypergraph convolutional networks (HCNs) to facilitate intra-modal and inter-modal interactions between multimodal graphs | Cross-validation | - | 5 | - | - | - | Yes | Yes | Yes | Yes |
| Ji, 2024 (32) | **HGRS**, (Renal ) CCRCC | - | 20x, 40x | 1000x1000 | Random | 20 | Hand crafted | CellProfiler | Overlap of SVM-RFE and LASSO-COX | - | WGCNA | Linear regression (top module), top hub genes of module | No | Cross-validation | 70:30 | 5 | - | - | - | - | - | Yes | - |
| Li, 2020 (33) | **DeepHit**, (Breast) - | Augmentation cropping, orientation, colour | - | 256x256 | - | - | Learned | ResNet - tissue heat map, DCGMM (CNN) - nuclear | LSTM feature encoder | Filter invalid data, normalise and standardise | - | Gene encoder | No | Cross-validation | - | 5 | - | - | - | - | - | Yes | - |
| Li, 2021 (34) | **HGPF**, (Colon) Adenocarcinoma | - | - | 1000x1000 | Random | 20 | Hand crafted | CellProfiler | Overlap of SVM-RFE and LASSO-COX | Normalised | WGCNA | Gene module most associated pathology features | No | Cross-validation | 70:30 | 10 | - | - | - | - | - | - | - |
| Li, 2022 (35) | **HFBSurv**, (Multi -10) Breast | - | - | 1000x1000 | - | 10 | Hand crafted | CellProfiler | RandomForest SRC | missing values removed, discretised gene expression and CAN. | - | RandomForestSRC | Modality-specific and cross-modality attentional factorized bilinear modules | Cross-validation | 80:20 | 5 | - | - | - | - | Yes | - | Yes |
| Liu, 2023 (36) | **MGCT**, (Multi - 5) All | CLAM tissue segmentation | 20x | 256x256 | - | - | Learned | ResNet50 (ImageNet) | All | - | SNN | All | Mutual-guided cross-modality attention mechanism | Cross-validation | 80:20 | 5 (Monte Carlo) | - | - | - | - | Yes | - | Yes |
| Liu, 2024 (37) | **IntraSA-InterCA**, (Breast) - | - | 40x | 1000x1000 | Highest image density patches per WSI | 10 | Hand crafted | CellProfiler | Fselector | Categorise into under, over or baseline expression | - | Fselector | Intra- and inter-modality attention mechanisms | Cross-validation | 80:20 | 5 | - | - | - | - | - | Yes | - |
| Lv, 2021 (38) | **PG-TFNet**, (Colorectum) - | - | 1x, 5x, 20x | 512x512 | Expert ROI annotation | - | Learned | ResNet based tribranch module for features at each scale, then transformer-based fusion for WSI vector | All | - | - | Differential gene expression by negative binomial distribution (DESeq) | Cross-modality transformer fusion | Cross-validation | 90:10 | 10 | - | - | - | - | - | - | - |
| Lv, 2023 (39) | **Trans-Surv**, (Colorectum) - | - | Thumbnail, 5x, 20x | 392x392 (5x), 224x224 (20x) | Random sampling of expert annotated ROI | - | Learned | Multiple CNNs (muttiscale + Patch tissue classifier) | All | CNV estimated by GISTIC2 method and then categories in 1 of 5 groups | - | Overlap of DESeq, edgeR, limma selection | Cross-attention transformer fusion | Cross-validation | 80:20 | 5 | - | - | - | - | - | - | - |
| Ning, 2020 (40) | **Gene+His***, (Renal) CCRCC | Stain normalisation | - | 128x128 | - | 150 (av.) | Learned | CNN | Block filtering post-pruning search (BFPS) algorithm | Transformed from read counts to normalised fragments per kilobase million. | Selection of most variant 8000 (to reduce noise), then WGCNA on these genes | Block filtering post-pruning search (BFPS) algorithm | Cross-attention fusion model | Cross-validation | 67:33 | 10 | - | - | - | Yes | - | - | - |
| *Ning, 2023 (41)* | **McLR Framework**, *(Multi - 3 ) LIHC* | *Stain normalisation* | *-* | *512x512* | *Eliminate if <50% tissue coverage* | *-* | *Hand crafted* | *Parameter free threshold adjacency statistics* | *PCA* | *-* | *-* | *PCA* | Ranking and regression constraints for cross-modal learning | *Cross-validation* | *-* | *10* | *-* | *-* | *-* | *-* | *Yes* | *Yes* | *Yes* |
| *Perez-Herrera, 2024 (42)* | **Model 3**, *(Breast) -* | *Stain normalisation (Reinhard method)* | *-* | *512x512* | *Expert annotations* | *-* | *Learned* | *UNet-VGG16 to predict percentage of defined tissue classes* | *% tumour, stroma, necrosis* | *Transformed* | *Stochastic gradient descent classifier (SNP), AdaBoost (expression), Decision tree (CNV)* | *Top 100 frequently mutated genes (CNV and CNP), top 30 known differentialy expressed genes in BC* | No | *Unclear* | *90:10* | *-* | *-* | *-* | *-* | *-* | | *-* | *Yes* |
| *Qiu, 2024 (43)* | **DDM-net**, *(Brain) Glioma* | *Otsu algorithm for tissue masks* | *5x, 10x, 20x* | *512x512* | *Patches that contain <30% tissue discarded* | *-* | *Learned* | *ResNet50 (ImageNet) with patch level clustering into 10 phenotypes using k-means (300x1024 dimensional embeddings) condensed by pathology encoders and concatenated to single multiscale patient level feature* | *-* | *-* | *SNN-based genomic encoder* | *Top 5000 variables with highest variance each omic modality. SVM-RFE selects 162 features fed into SNN* | Achieved through dual-space disentanglement via separate variational autoencoders (VAEs), enabling the model to learn shared and modality-specific representations | *Cross-validation* | *-* | *5* | *-* | *-* | *-* | *-* | *-* | *Yes* | *-* |
| *Shao, 2020 (44)* | **OMMFS,** *(Multi - 3) LUSC* | *-* | *-* | *5000x5000* | *Regions defined by experts* | *4-6* | *Hand crafted* | *Extract 10 types nuclear features per patch. Then 10 bin histogram using 5 statistics* | *Log rank, OMMFS* | *-* | *co-expression network analysis, then summarise gene modules by singular value docomposition* | *Log rank test, OMMFS* | Via Ordinal Multi-modal Feature Selection (OMMFS) | *Random split data* | *80:20* | *N/A* | *-* | *-* | *-* | *-* | *Yes* | *-* | *Yes* |
| *Shao, 2023 (45)* | **FAM3L,** *(Multi - 3) Multi (KIRP)* | *-* | *-* | *1024x1024* | *Image density >0.7 (based on percentage of non-white values)* | *-* | *Hand crafted* | *Unsupervised nuclear segmentation, cell level morphological and topological features extracted. Then 10 bin histogram with 5 statistics used to aggreage to image level* | *-* | *-* | *lmQCM* | *All* | Via Hilbert-Schmidt Independence Criterion (HSIC) | *Cross-validation* | *-* | *5* | *-* | *-* | *-* | *-* | *Yes* | *-* | *Yes* |
| *Shao, 2020 (46)* | **M2DP,** *(Multi - 3) BRCA* | *-* | *-* | *3000x3000* | *ROIs chosen by expert* | *2-8* | *Hand crafted* | *Unclear* | *Unclear - 7 cell level features from extracted nuclei aggregated into 105 dimension feature* | *-* | *lmQCM* | *All* | Via Task Relationship Learning | *Cross-validation* | *-* | *5* | *-* | *-* | *-* | *-* | *Yes* | *Yes* | *Yes* |
| *Shao, 2023 (47)* | **IMO-TILS,** *(Multi - 3) Breast* | *-* | *-* | *512x512* | *1) By RGB densitity 2)200 patches with largest TIL ratio* | *2-8* | *Learned* | *Unet++ (segment tILS and tumour tissue), ResNet-101, KNN* | *Graph attention network pooling* | *-* | *lmQCM* | *Concrete autoencoder (miRNa and mRNA)* | Achieved through a deep generalized canonical correlation analysis (DGCCA) with an attention mechanism | *Cross-validation* | *80:20* | *5* | *-* | *-* | *-* | *-* | *Yes* | *Yes* | *Yes* |
| *Steyaert, 2023 (48)* | **Late fusion**, *(Brain - Multi) Adult Glioblastoma* | *Stain augmentation, otsu tissue segmentation* | *20* | *224x224* | *Random* | *100* | *Learned* | *ResNet50 CNN* | *All* | *Missing values removed, Gene counts normalised by log and z score transformation, batch effect adjustment (Combat-Seq)* | *Multi-layer perceptron* | *-* | No | *Cross-validation* | *80:20* | *10* | *0.778* | *x* | *Yes* | *-* | *Yes* | *Yes* | *Yes* |
| *Subramanian, 2021 (49)* | **GCCA,** *(Breast) -* | *Nuclei segmentations from another study* | *-* | *2000 x2000* | *Random* | *25* | *Hand crafted* | *CellProfiler* | *Averaged across patches* | *-* | *-* | *Selection of most variant genes and correspinding z-scores using coefficient of variation of log transformed expression values,then use of STRING database to capture prior knowledge and assign* | Via sparse canonical correlation analysis (SCCA) to capture both intra-modality and inter-modality correlations | *Cross-validation* | *70:30* | *5* | *-* | *-* | *Yes* | *-* | *-* | *Yes* | *-* |
| *Subramanian, 2024 (50)* | **SCCA**, *(Breast) -* | *Nuclei segmentation masks generated* | *-* | *2000 x2000* | *Random* | *25* | *Hand crafted* | *CellProfiler* | *5 bin histgram for each of the 215 features* | *-* | *-* | *Most variant genes using coefficient variation of log2 transformed expression values, selecting top 1000 genes.* | Via a probabilistic graphical model | *Cross-validation* | *75:25* | *5* | *-* | *-* | *Yes* | *-* | *-* | *Yes* | *-* |
| *Sun, 2018 (51)* | **GPMKL**, *(Breast) -* | *-* | *40x* | *1000x1000* | *Densest tiles per WSI* | *10* | *Hand crafted* | *CellProfiler* | *Fselector* | *Gene expression normalised and discretised into under, over or baseline expression. Gene methy.ation and protein expression normalised to z-score.* | *-* | *F-selector (information gain ratio measure)* | Using multiple kernel learning techniques | *Cross-validation* | *80:20* | *10* | *-* | *-* | *-* | *-* | *-* | *Yes* | *-* |
| *Tan, 2022 (52)* | **MultiCoFusion**, *(Brain) GBMLGG* | *-* | *-* | *512x512* | *Cropped from 1024x1024 expert chosen ROI* | *-* | *Learned* | *ResNet-152 (ImageNet)* | *All* | *-* | *Sparse graph convultional neural network (SGCN)* | *-* | Through multi-task correlation learning | *Cross-validation* | *80:20* | *15 randomised assignme* | *-* | *-* | *Yes* | *-* | *-* | *Yes* | *-* |
| Vale-Silva, 2021 (53) | **MultiSurv**, (Multi - 33) Multi | Otsu algorthm for tissue segmentation, Orientation and colour | - | 299x299 | - | - | Learned | ResNet50 CNN (ImageNet) | All | CNV categorised into loss, gain or neutral, batch normalisation for omic and clinical modalities | FC CNN | All | No | Bootstrap | 90:10 | ? | - | - | - | Yes | Yes | Yes | Yes |
| Vollmer, 2024 (54) | **Random Survival Forest**, (Oral) SCC | Colour normalization | 40x | 1024x1024 | 10 Highest density patches (ref 19) | - | Hand crafted | CellProfiler | - | Gene filtering for missing values and data normalisation steps to adjust for variability in the data | Top 200 diffferentially expressed | PCA and ElasticNet | No | Cross-validation | - | 5 | - | - | Yes | - | - | Yes | - |
| Wang, 2021 (55) | **GPDBN**, (Breast) - | - | - | 1000x1000 | 10 Highest density patches (ref 19) | - | Hand crafted | CellProfiler | Fselector | Normalisation with Z score, discretised into under, over, baseline. | - | Fselector | Bilinear pooling, enabling interaction between image and genomic features | Cross-validation | 80:20 | 5 | - | - | Yes | - | - | Yes | - |
| Wang, 2023 (56) | **HC-MAE**, (Multi - 6 ) LGG | CLAM tissue segmentation | - | 16x16, 256x256, 4096x4096 | If >1 WSI per patient - randomly select 1. CLAM to select sub-regions with high diagnostic value | - | Learned | pre-trained Vision Transformer-based Masked Autoencoders | All | Remove genes with zero variance, differential gene expression analysis | MLP | Remove genes with zero variance, differential gene expression analysis, then Random survival forest to select based on feature importance | Hierarchical cross-attention mechanisms integrating histopathological image representations and multi-omics data | Cross-validation | - | 5 | - | - | - | - | Yes | Yes | Yes |
| Wei, 2023 (57) | **MultiDeepCox-SC**, (Stomach) - | Colour normalisation, also data augmentation (orientation) | s selection of c | 512x512 | 1/9 512x512 from larger 1536x1536 patch (chosen from whole image due to high nuclei density) | - | Hand crafted | CellProfiler | SIS + LASSO | - | Removed or selected based on variance | Intersection of values selected by Sure independent screening + LASSO | No | Cross-validation | - | 10 | - | - | - | Yes | - | Yes | - |
| Wu, 2023 (58) | **CAMR**, (Multi - 3) LGG | - | - | 1000x1000 | Selected by highest RGB image density | 10 | Hand crafted | CellProfiler | RandomForest SRC | Categorisation into under, over, baseline expression | - | RandomForestSRC | Adversarial alignment and cross-modality fusion modules | Cross-validation | 80:20 | 5 | - | - | - | - | Yes | - | Yes |
| Xie, 2024 (59) | **GaCaMML**, (Stomach) | Tissue segmentation | 40x | 256x256 | - | >10,000 | Learned | ResNet50 (ImageNet) | All | - | Two layer FC CNN with ELU activation function | 231 genes from 9 chosen oncology pathways | Cross-modal attention mechanism that aligns and fuses features from histopathological images and gene expression data. Additionally, Multiple Instance Learning (MIL) is applied to handle the variable sizes of the image patches and to learn spatially and contextually relevant information from them. | Cross-validation | 80:20 | 5 | 0.573 | x | Yes | - | - | - | - |
| Zeng, 2020 (60) | **Multi-omics model**, (Head & Neck) SCC | Tissue segmentation | - | 1000x1000 | Random | 20 | Hand crafted | CellProfiler | All | Normalisation transcriptomic data | - | 100 most common somatic mutation, 100 most DEG (transcriptoimic), All | No | Cross-validation | 50:50 | 10 | - | - | - | Yes | - | Yes | - |
| Zeng, 2021 (61) | **Multi-omics model**, (Ovary) HGSOC | - | - | 1000x1000 | Random | 60 | Hand crafted | CellProfiler | All | - | - | 100 most common somatic mutation, 100 most DEG, all | No | Cross-validation | 50:50 | 5 | - | - | - | Yes | - | Yes | - |
| Zhan, 2021 (62) | **Cox-nnet**, (Liver) HCC | Patches processed to rescale RGB values to selected slide, tissue folds removed | - | 1000x1000 | Expert annotated ROI, 10 densest patches within region by RGB | 10 | Hand crafted | CellProfiler | All | Normalised into RPKM by TCGA-Assembler | - | - | Two-stage Cox-nnet model that integrates hidden features from separate Cox-nnet models for histopathology and transcriptomics | Cross-validation | 80:20 | 5 | - | - | - | - | - | Yes | - |
| Zhang, 2020 (63) | **HI-MKL**, (Brain) Glioblastoma | - | 20 | 1024x1024 | Highest nuclear density | 20 | Hand crafted | CellProfiler | Mutual feature selection methods (mRMR) | Normalised by calculating z index of all variables | - | Mutual feature selection methods (mRMR) | Using multiple kernel learning techniques | Cross-validation | - | 10 | - | - | Yes | - | - | Yes | - |
| Zhao, 2023 (64) | **Ada-RSIS**, (Multi - 3 ) UCEC | - | - | - | - | - | Hand crafted | Parameter-free threshold adjacency statistics | Principal component analysis | - | - | Principal component analysis | Joint learning of sharable and individual subspaces from multi-modality data, incorporating intra-modality complementarity and inter-modality incoherence. | Cross-validation | - | 10 | 0.672 | x | - | - | Yes | Yes | Yes |
| Zheng, 2024 (65) | **FSM**, (Lung - Multi) LUAD | Tissue segmentation | - | - | - | - | Learned | CNN (pretrained) | - | Normalisation of gene counts (EdgeR Biconductor package), filtering out of duplicate and invariant genes, batch correction (ComBat) | FC CNN | Gene signatures assiciated with B cell populations, | Graph attention mechanism that integrates embeddings from pathology images and gene expression data | Cross-validation | - | 5 | 0.579 (c stat) | 0.006 (StD) | - | - | - | Yes | - |
| Zhou, 2023 (66) | **CMTA**, (Multi - 5) GBMLGG | CLAM tissue segmentation | 40 | 512x512 | - | - | Learned | ResNet-50 (ImageNet) | - | - | Grouped into functional 6 oncology categories then fed in to FC CNN | - | Cross-modal attention module that facilitates interactions between pathology images and genomic profiles | Cross-validation | - | 5 | - | - | - | - | Yes | - | Yes |
| Zhou, 2023 (67) | (Colon) Adenocarcinoma | Colour normalisation (Macenko), backgroud removed | 20 | 512x512 | >50% overlap with expert generated tumour annotation mask | - | Learned | Inception V3 (ImageNet) | All | - | - | 207 genes from 11 canonical pathways, filtered to remove genes not frequently mutated in TCGA-COAD | No | Cross-validation | 70:30 | 100 (Monte carlo) | Unclear | - | - | - | - | Yes | - |
| Zhou, 2024 (68) | **MSEN**, (Multi - 5 ) GBM-LGG | - | - | 4096x4096 | Removal of patches not meeting crieria for HIPT model | - | Learned | HIPT (Hierarchical image pyramid transformer) | All | - | - | Copy number frequency, mutation frequency >10%/5% repsectively. Genes with relevance by Molecular Sigantures Database categorised into 6 functional gene classes. | Cross-modality attention mechanism | Cross-validation | 80:20 | 5 | - | - | - | - | Yes | Yes | Yes |
| Zhu, 2023 (69) | **SAMMS**, (Multi - 2) LGG | ScoreTiler tool for image segmentation | 20 | 4096x4096 | - | - | Learned | SAM (pretrained) | ElasticNet (most salient across both modalities) | z score normalisation across datasets | - | 2000 Features with most significant variance across modalities, then ElasticNet (most salient across both modalities) | Achieved through a foundation model that learns joint representations across modalities | Cross-validation | - | 5 | - | - | - | Yes | Yes | Yes | Yes |

**Figure S1. Modelling method trends over time**

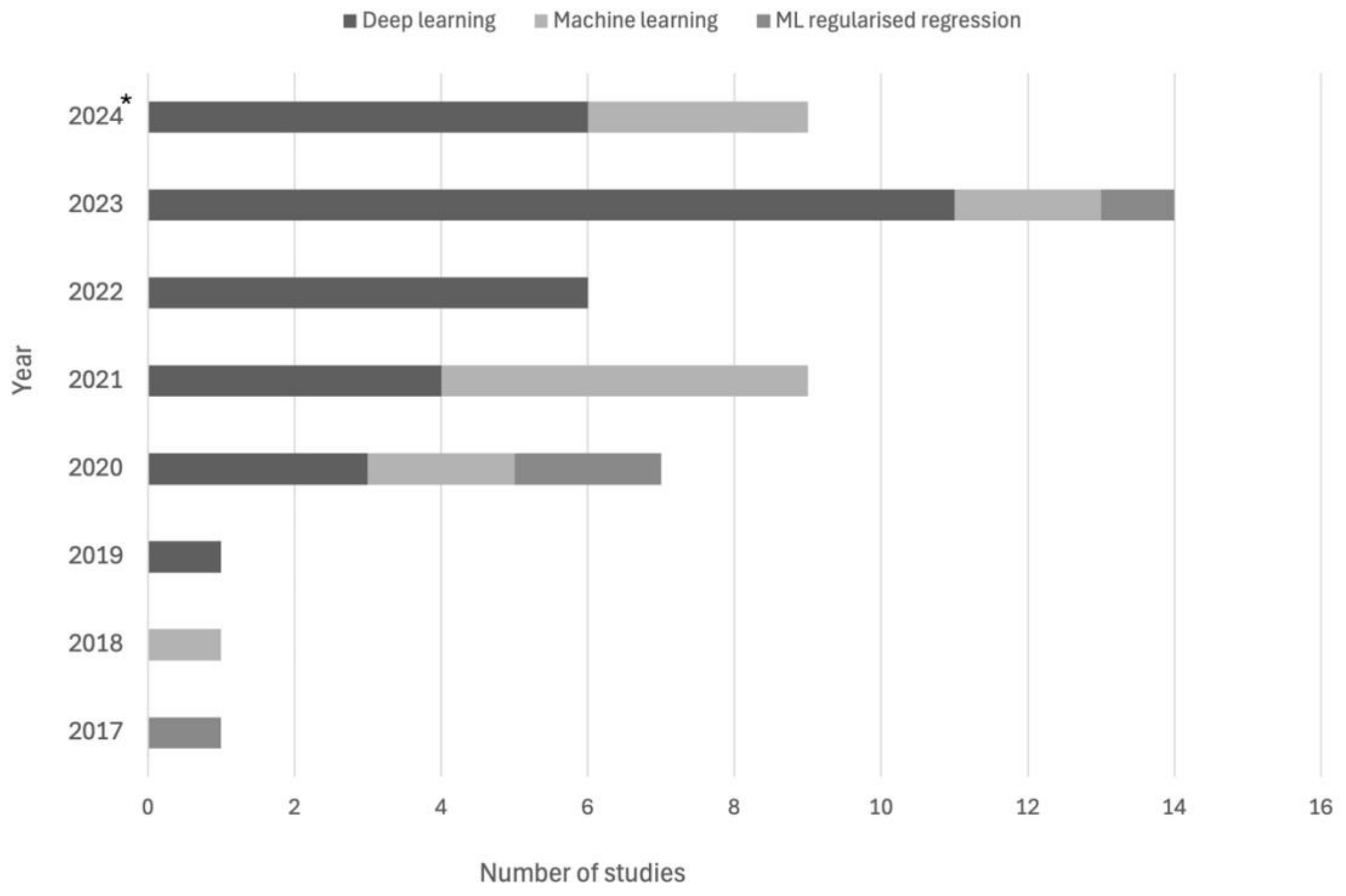

**Figure S1. Prediction modelling trends.**

This chart highlights the use of different modelling approaches used by the studies of this review, and presents their use over time. *Searches conducted up to August 2024. The apparent drop in overall studies for 2024 is most likely to the mid-year date of the searches. Studies were assigned to one of three approaches: ML regularised regression, classical machine learning or deep learning. Categories used are defined in the supplementary section "author definitions". There is a predominance of deep learning methods in studies of this review, increasing rapidly in number since 2019.

### 1.11.2 Figure S2. Model performance by cancer dataset

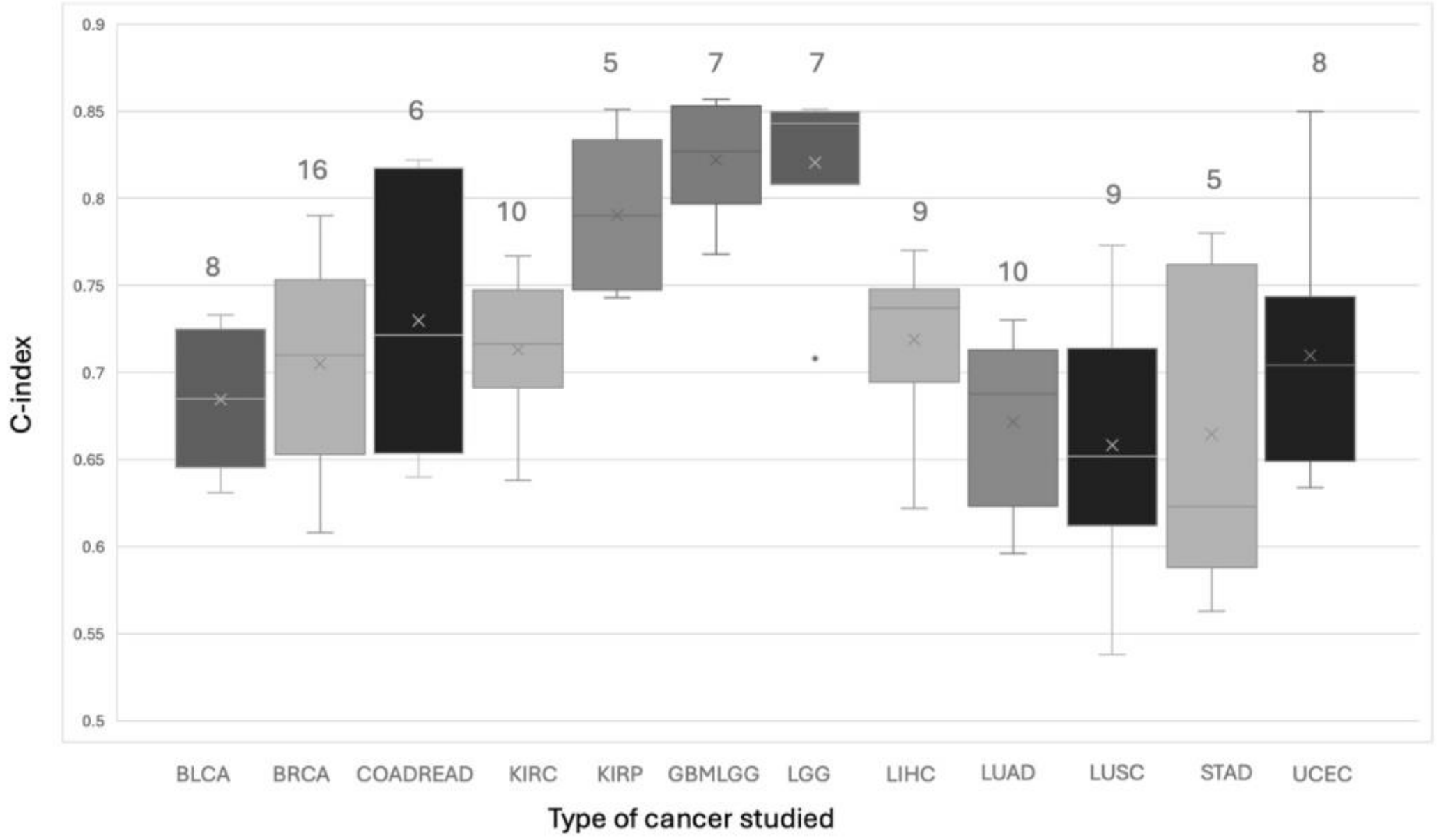

**Figure S2. Model performance across cancer types**

This comparison was performed for cancer types in which there were more five or more results across the studies in this review. Results from alternate datasets to the one defined as the model of interest for the purposes of this review are also included in this analysis. The number of results included is indicated above the box and whisker plot for each cancer type. Cancer types use the abbreviation nomenclature used by The Cancer Genome Atlas project; BLCA (bladder, urothelial carcinoma)[23,26,35,36,64,66,68,69], BRCA (breast, invasive carcinoma)[23,26,33,35-37,45-47,50,51,55,56,58,66,68], COADREAD (colon and rectum, adenocarcinoma)[23,26,35,38,39,56], KIRC (kidney, renal clear cell carcinoma)[23,26,27,31,35,40,41,44,45,47], KIRP (kidney, renal papillary cell carcinoma)[23,26,35,44,45], GBMLGG (brain, glioblastoma multiforme and lower grade glioma)[27,36,43,48,52,66,68], LGG (brain, lower grade glioma)[23,26,35,56,58,64,69], LIHC (liver, hepatocellular carcinoma)[23,26,30,31,35,41,46,56,62], LUAD (lung, adenocarcinoma)[23,26,31,35,36,41,56,65,66,68], LUSC (lung, squamous cell carcinoma)[23,26,31,35,44,46,47,58,65], STAD (stomach, adenocarcinoma)[23,26,56,57,59], UCEC (Uterine corpus, endometrial carcinoma)[23,26,31,35,36,64,66,68]. Reference numbers link to the reference list of the main document.

Performance is shown to be highest in brain tumours (glioblastoma and low grade glioma) and in renal papillary cell carcinoma. Statistical analysis was unsuitable for these heterogenous studies and this should be regarded as a descriptive comparison only.

### 1.11.3 Figure S3. Model performance by sample size

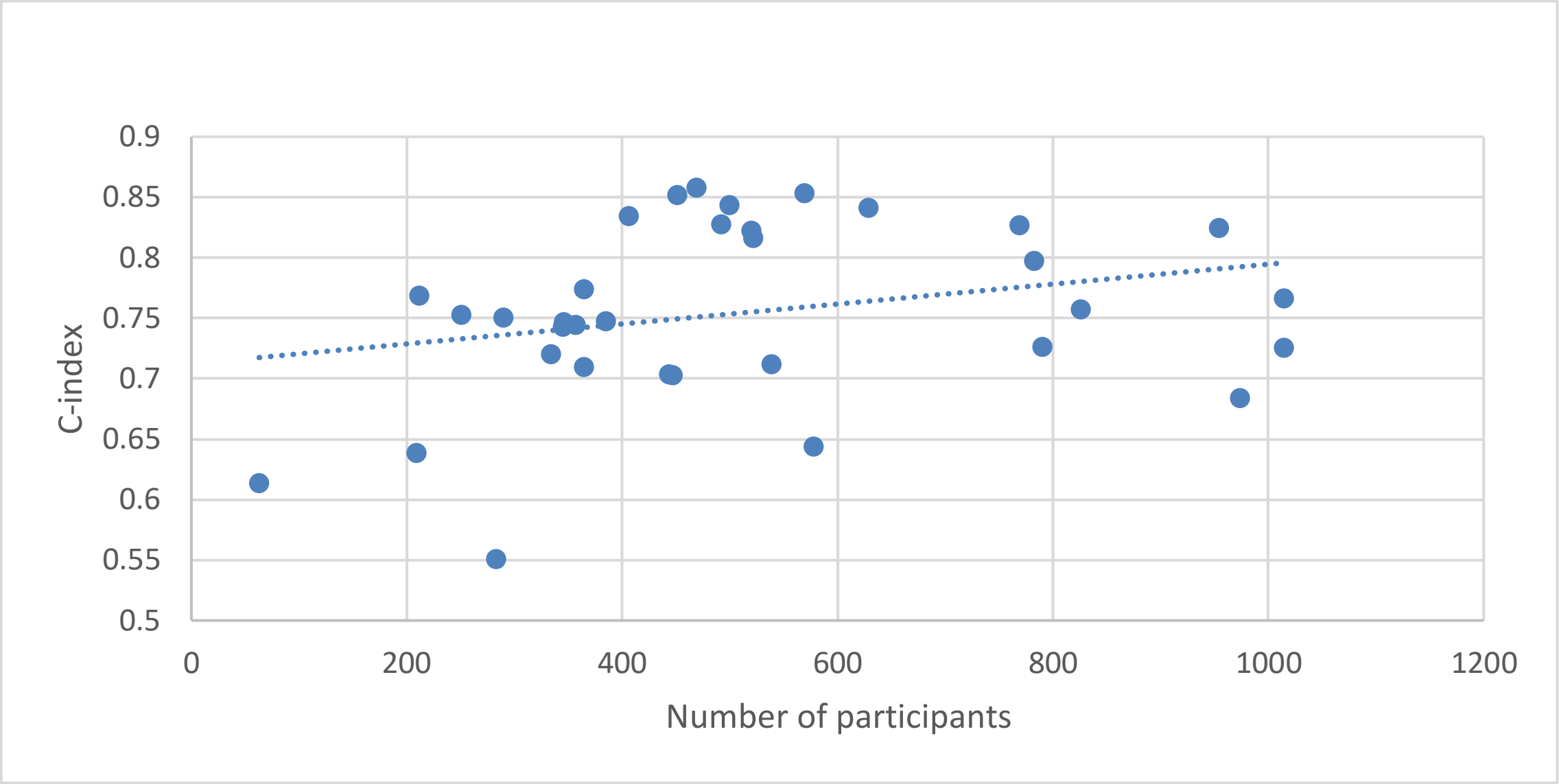

**Figure S3. Model performance in relation to participant number.** Multimodal model performance of the model of interest compared with number of participants. Thirty nine studies presented results by c-index, however in one study the sample size was unclear, so 38 studies were eligible for inclusion in this comparison. The total number of participants in a study may be greater than presented here due to studies which evaluated their model on multiple datasets separately. There is a trend towards improved performance with increased participant number but studies with sample sizes of ≥400 appear to be sufficient for optimal performance.

### 1.11.4 Figure S4. Model performance by event rate

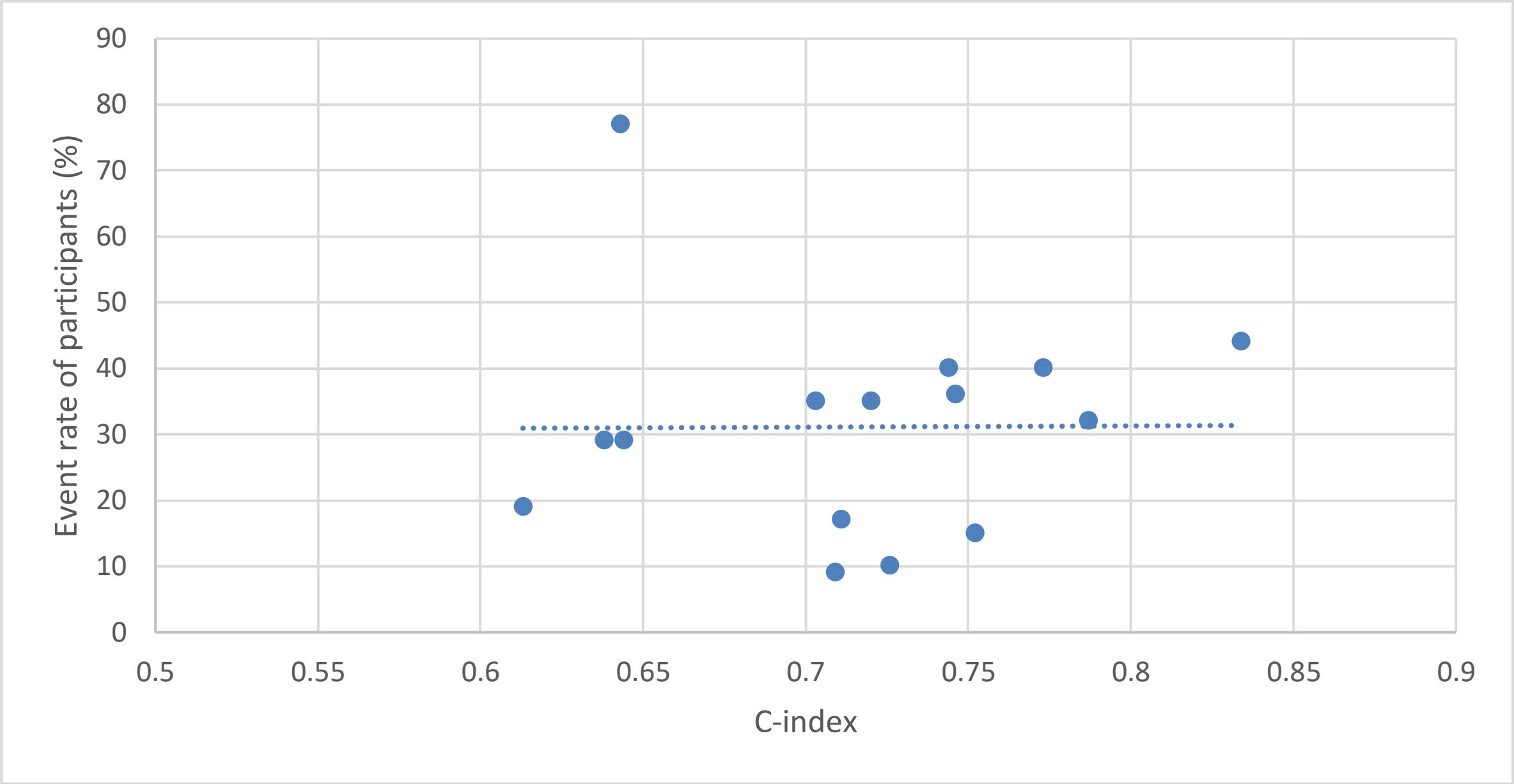

**Figure S4.**

The relationship between outcome event rate (y-axis) in the studies and performance measured by c-index (x-axis) was explored. Only 15 studies, were eligible for evaluation due to missing values within other studies. There is no apparent relationship between event rate and reported performance in this limited assessment, as demonstrated by the trend line.